\documentclass[sigconf,screen]{acmart}

\setcopyright{acmcopyright}
\acmPrice{15.00}
\acmDOI{10.1145/3533767.3534388}
\acmYear{2022}
\copyrightyear{2022}
\acmSubmissionID{issta22main-p189-p}
\acmISBN{978-1-4503-9379-9/22/07}
\acmConference[ISSTA '22]{Proceedings of the 31st ACM SIGSOFT International Symposium on Software Testing and Analysis}{July 18--22, 2022}{Virtual, South Korea}
\acmBooktitle{Proceedings of the 31st ACM SIGSOFT International Symposium on Software Testing and Analysis (ISSTA '22), July 18--22, 2022, Virtual, South Korea}

\usepackage{graphicx}
\usepackage[many]{tcolorbox}

\usepackage{mathtools,amssymb,latexsym,amsfonts,stmaryrd}
\usepackage{pbox}
\usepackage{xfrac}
\usepackage{booktabs}
\usepackage{xcolor}
\usepackage{threeparttable}
\usepackage{amsthm}
\usepackage{hyperref}
\usepackage{multirow}
\usepackage{tikz}
\usepackage{mathrsfs}
\usepackage{mathpartir}
\usepackage{algorithm}
\usepackage{url}
\usepackage{syntax}
\usepackage{framed}
\usepackage[noend]{algpseudocode}
\usepackage{flushend}
\usepackage{listings}
\usepackage{moresize}
\usepackage{wrapfig}
 
\usepackage{seqsplit}
\usepackage{alltt}
\usepackage{xspace}
\usepackage{pgfplots}
\usepackage{subcaption}
\usetikzlibrary{positioning}
\usepackage{enumitem}
\usepackage{pifont}% http://ctan.org/pkg/pifont

\DeclareMathAlphabet{\mathcal}{OMS}{cmsy}{m}{n}

\usepackage{amsmath, listings, amsthm, amssymb, proof, xspace}

\usepackage{thmtools}

\declaretheoremstyle[spaceabove=\topsep,notefont=\normalfont\itshape]{mystyle}

\newcommand{\revise}[2]{{\color{red}{\ifx&#1&\else- #1\fi}} {\color{ForestGreen}{\ifx&#2&\else+ #2\fi}}}%
\renewcommand{\revise}[2]{#2}%

\usetikzlibrary{arrows,patterns, decorations.pathreplacing}
\newtheorem{definition}{Definition}

\newcommand{\F}{Fig.}
\newcommand{\T}{Table}
\renewcommand{\S}{Sec.}
\newcommand{\A}{Alg.}
\newcommand{\E}{Eq.}

\newcommand{\ignore}[1]{}

\lstdefinestyle{base}{
  moredelim=**[is][\color{red}]{@}{@},
  escapeinside={<@}{@>}
}

\newcommand{\tool}{\textsc{MDPFuzz}\xspace}

\usepackage{amssymb}% http://ctan.org/pkg/amssymb
\usepackage{pifont}% http://ctan.org/pkg/pifont

\newcommand\DejaVuttfamily{%
  \fontfamily{DejaVuSansMono-TLF}\selectfont }

\lstdefinestyle{base}{
  moredelim=**[is][\color{red}]{@}{@},
  escapeinside={<@}{@>}
}

\lstdefinelanguage
   [x64]{Assembler}     % add a "x64" dialect of Assembler
   [x86masm]{Assembler} % based on the "x86masm" dialect
   {morekeywords={CDQE,CQO,CMPSQ,CMPXCHG16B,JRCXZ,LODSQ,MOVSXD, %
                  POPFQ,PUSHFQ,SCASQ,STOSQ,IRETQ,RDTSCP,SWAPGS, %
                  rax,rdx,rcx,rbx,rsi,rdi,rsp,rbp, %
                  r8,r8d,r8w,r8b,r9,r9d,r9w,r9b}} % etc.

\usepackage{listings}
\usepackage{fancyvrb}
\usepackage{color}
\definecolor{lightgray}{rgb}{.9,.9,.9}
\definecolor{darkgray}{rgb}{.4,.4,.4}
\definecolor{purple}{rgb}{0.65, 0.12, 0.82}
\definecolor{commentgreen}{RGB}{63,127,95}

\colorlet{myPurple}{blue!40!red}
\definecolor{myOrange}{RGB}{255,192,0}
\newcommand{\code}[1]{\textcolor{blue}{\textsc{\textit{\bfseries{#1}}}}}

\lstdefinelanguage{Solidity}{
  keywords={len,delete,int,void,payable, public, event, contract, typeof, new, true, false, catch, function, return, null, catch, switch, var, if, in, while, do, else, case, break,struct,const,socklen_t,sa_familty_t,char,sockaddr},
  keywordstyle=\color{violet}\bfseries,
  ndkeywords={class, export, boolean, throw, implements, import, this},
  ndkeywordstyle=\color{darkgray}\bfseries,
  identifierstyle=\color{black},
  sensitive=false,
  comment=[l]{//},
  escapeinside={(*@}{@*)},          % if you want to add LaTeX within your code
  morecomment=[s]{/*}{*/},
  commentstyle=\color{commentgreen}\ttfamily,
  stringstyle=\color{red}\ttfamily,
  morestring=[b]',
  morestring=[b]"
}

\newcommand{\rnum}[1]{\uppercase\expandafter{\romannumeral #1\relax}}

\usepackage{xcolor}

\algnewcommand{\LeftComment}[1]{\Statex \(\triangleright\) #1}

\definecolor{pptbrown}{RGB}{132,60,12}
\definecolor{pptgreen}{RGB}{56,87,35}

\definecolor{pptgrey}{RGB}{202,202,202}
\definecolor{pptgreen1}{RGB}{192,217,170}

\let\OLDthebibliography\thebibliography
\renewcommand\thebibliography[1]{
  \OLDthebibliography{#1}
  \setlength{\parskip}{0pt}
  \setlength{\itemsep}{0pt plus 0.1ex}
}

  \DeclareFontFamily{U}{dutchcal}{\skewchar \font =45}
  \DeclareFontShape{U}{dutchcal}{m}{n}{
    <-> dutchcal-r}{}
  \DeclareFontShape{U}{dutchcal}{b}{n}{
    <-> dutchcal-b}{}
  \DeclareMathAlphabet{\mdutchcal}{U}{dutchcal}{m}{n}
  \SetMathAlphabet{\mdutchcal}{bold}{U}{dutchcal}{b}{n}
  \DeclareMathAlphabet{\mdutchbcal} {U}{dutchcal}{b}{n}

  \DeclareFontFamily{U}{txcal}{\skewchar \font =45}
  \DeclareFontShape{U}{txcal}{m}{n}{
    <-> txr-cal}{}
  \DeclareFontShape{U}{txcal}{b}{n}{
    <-> txb-cal}{}
  \DeclareMathAlphabet{\mtxcal}{U}{txcal}{m}{n}
  \SetMathAlphabet{\mtxcal}{bold}{U}{txcal}{b}{n}
  \DeclareMathAlphabet{\mtxbcal} {U}{txcal}{b}{n}

\renewcommand{\algorithmiccomment}[1]{\hfill$\triangleright$\textit{#1}}
\newlength{\textfloatsepsave}
\newcommand\freefootnote[1]{%
  \let\thefootnote\relax%
  \footnotetext{#1}%
  \let\thefootnote\svthefootnote%
}

\begin{document}

\title{\tool: Testing Models Solving Markov Decision Processes}

\author{Qi Pang}
\affiliation{%
  \institution{The Hong Kong University of Science and Technology}
  \country{Hong Kong, China}
}
\email{qpangaa@cse.ust.hk}

\author{Yuanyuan Yuan}
\affiliation{%
  \institution{The Hong Kong University of Science and Technology}
  \country{Hong Kong, China}
}
\email{yyuanaq@cse.ust.hk}

\author{Shuai Wang}
 \affiliation{%
  \institution{The Hong Kong University of Science and Technology}
  \country{Hong Kong, China}
 }
\authornote{Corresponding Author}
\email{shuaiw@cse.ust.hk}

\begin{abstract}

 The Markov decision process (MDP) provides a mathematical framework for
 modeling sequential decision-making problems, many of which are crucial to
 security and safety, such as autonomous driving and robot control. The rapid
 development of artificial intelligence research has created
 efficient methods for solving MDPs, such as deep neural networks (DNNs),
 reinforcement learning (RL), and imitation learning (IL). However, these
 popular models solving MDPs are neither thoroughly tested nor rigorously
 reliable.

 We present \tool, the first blackbox fuzz testing framework for models solving
 MDPs. \tool\ forms testing oracles by checking whether the target model enters
 abnormal and dangerous states. During fuzzing, \tool\ decides which mutated
 state to retain by measuring if it can reduce cumulative rewards or
 form a new state sequence. We design efficient techniques to quantify the
 ``freshness'' of a state sequence using Gaussian mixture models (GMMs) and
 dynamic expectation-maximization (DynEM). We also prioritize states with
 high potential of revealing crashes by estimating the local sensitivity of target
 models over states.

 \tool\ is evaluated on five state-of-the-art models for solving MDPs, including supervised DNN, RL, IL, and multi-agent RL. Our evaluation
 includes scenarios of autonomous driving, aircraft collision avoidance, and
 two games that are often used to benchmark RL. During a 12-hour run, we find
 over 80 crash-triggering state sequences on each model. We show inspiring findings that
 crash-triggering states, though they look normal, induce distinct neuron activation
 patterns compared with normal states. We further develop an abnormal 
 behavior detector to harden all the evaluated models and repair them with the
 findings of \tool\ to significantly enhance their robustness without
 sacrificing accuracy.

\end{abstract}




\maketitle

\section{Introduction}
\label{sec:introduciton}

Recent advances in artificial intelligence (AI) have improved our ability to solve
decision-making problems by modeling them as Markov decision processes
(MDPs). Modern learning-based solutions, such as deep neural networks (DNNs),
reinforcement learning (RL), and imitation learning (IL), tackle decision-making
problems by using the inherent properties of MDPs. These solutions have already
demonstrated superhuman performance in video games~\cite{berner2019dota},
Go~\cite{silver2016mastering}, robot control~\cite{kober2013reinforcement},
and are being deployed in mission-critical scenarios such as collision avoidance
and autonomous driving~\cite{julian2019deep,BADUE2021113816,apollo}. The
well-known Aircraft Collision Avoidance System X (ACAS
Xu)~\cite{marston2015acas} employs a search table to model the airplane's policy in an MDP. 
Several DNN-based variants of ACAS Xu are also proposed and well-studied to further reduce the memory needed without sacrificing performance~\cite{julian2019deep,wang2018formal}. 
It predicts the optimal course of action based on the positions
and speeds of intruder planes. It has passed NASA and FAA tests~\cite{nasa,marston2015acas} and will
soon be installed in over 30,000 flights worldwide and the US Navy's fleets~\cite{NAVAIR,olson2015airborne,wang2018formal}. DNNs
and IL are also used by NVIDIA and Waymo (previously Google's self-driving car
project) to learn lane following and other urban driving policies using massive 
amounts of human driver
data~\cite{bansal2018chauffeurnet,bojarski2016end,bojarski2017explaining}.

Despite their effectiveness, these methods do not provide a
strict guarantee that no catastrophic failures will occur when they are used
to make judgments in real-world scenarios. Catastrophic failures are
intolerable, especially in security- or safety-critical scenarios. For
example, recently, a Tesla Model S collided with a fire vehicle at 65 mph while the
Autopilot system was in use~\cite{teslacrash}, and in 2016, Google's
self-driving car collided with the side of a bus~\cite{googlecrash}. In 2018, Uber's self-driving system experienced similar fatal errors~\cite{ubercrash}.

Software testing has been successfully deployed to improve the dependability of de
facto deep learning (DL) models such as image classification and object
detection models~\cite{pei2017deepxplore,zhang2018deeproad,zhang2020machine}. Existing studies, however,
are insufficient for testing models solving MDPs. 
From the \textbf{oracle's perspective}, previous DL testing methods often search for inconsistent model
predictions (e.g., via metamorphic testing or differential
testing~\cite{pei2017deepxplore}). However, as shown in \S~\ref{sec:motivation}, 
an ``inconsistent'' prediction seldom induces an abnormal and dangerous state (e.g., collisions
in autonomous driving) in a model solving MDPs. From the \textbf{input mutation's perspective}, previous DL tests typically mutate arbitrary model inputs (e.g.,
with a rain filter~\cite{zhang2018deeproad}) to stress the target model.
However, a typical model solving MDPs continuously responds to a sequence of states, and
modifying one or a few states is unrealistic due to the continuity of adjacent
states. From the \textbf{model complexity's perspective}, MDPs have complex and not differentiable 
state transitions, making it difficult, if not impossible, to develop an 
objective function to guide testing as is done in whitebox DL testing~\cite{pei2017deepxplore}. 
In addition, the models' internal states are unavailable in real-world blackbox settings.

This paper presents \tool, a blackbox fuzz testing framework for models solving
MDPs. \tool\ provides a viable and unified solution to the aforementioned
issues. First, \tool\ outlines a practical testing oracle for detecting the
models that enter severely \textit{abnormal states} (e.g., collisions in
autonomous driving). Second, instead of mutating arbitrary intermediate states,
\tool\ only mutates the initial state conservatively to ensure that the sequence
is realistic. The mutated initial states are also \textit{validated} in a
deliberate way to ensure realism. Third, \tool\ tackles blackbox scenarios,
allowing testing of (commercial) off-the-shelf models solving MDPs.
All of these factors, while necessary, add to the complexity and cost of testing
models solving MDPs. Therefore, \tool\ incorporates a series of optimizations.
\tool\ retains mutated initial states reducing cumulative rewards. It also
prefers mutated initial states that can induce ``fresh'' state sequences
(comparable to code coverage in software fuzzing). It measures the state
sequence freshness with constant and modest cost by leveraging MDP properties
and popular statistics mechanisms like Gaussian mixture models
(GMMs)~\cite{mclachlan1988mixture} and dynamic expectation-maximization (DynEM).
\tool\ also cleverly uses local sensitivity to assess an initial state's
potential to expose the models' new behavior (comparable to ``seed energy'' in
software fuzzing~\cite{bohme2017coverage}).

We evaluate five state-of-the-art (SOTA) models, including DNNs, deep RL,
multi-agent RL, and IL solving MDPs. The evaluated scenarios include games,
autonomous driving, and aircraft collision avoidance. \tool\ can efficiently
explore the state sequence space and uncover a total of 598 crash-triggering
state sequences for models we tested in a 12-hour run. Our findings show that
crash-triggering states, although considered natural and solvable by the tested
MDP environments, have distinct neuron activation patterns compared with normal
states across all examined models. We interpret that \tool\ can efficiently
cover the models' corner internal logics. Further, we rely on the uncovered
distinct neuron activation patterns to harden models, achieving a promising
detection accuracy of abnormal behaviors (over 0.78 AUC-ROC). We also repair the
models using the findings of \tool\ to notably increase their robustness
(eliminating 79\% of crashes) without sacrificing accuracy.
In summary, our contributions are as follows:

\begin{itemize}
    \item This paper, for the first time, proposes a general and effective fuzz
      testing framework particularly for models solving MDPs in blackbox
      settings.

    \item \tool\ makes several practical design considerations. To accelerate
      fuzzing and reduce the high testing cost, \tool\ incorporates a set of
      design principles and optimizations derived from properties of MDPs. 

    \item Our large-scale evaluation of five SOTA models subsumes different
      practical and security-critical scenarios. Under all conditions,
      \tool\ detects a substantial number of crashes-triggering state sequences.
      We further employ findings of \tool\ to develop an abnormal behaviors
      detector and also repair the models, making them far more resilient.
\end{itemize}

\begin{tcolorbox}[size=small]
We host the source code of \tool, data, and our findings at
\url{https://sites.google.com/view/mdpfuzz}.
\end{tcolorbox}
\section{Preliminary}
\label{sec:background}

\begin{figure}[!b]
    \centering
    \includegraphics[width=1.00\linewidth]{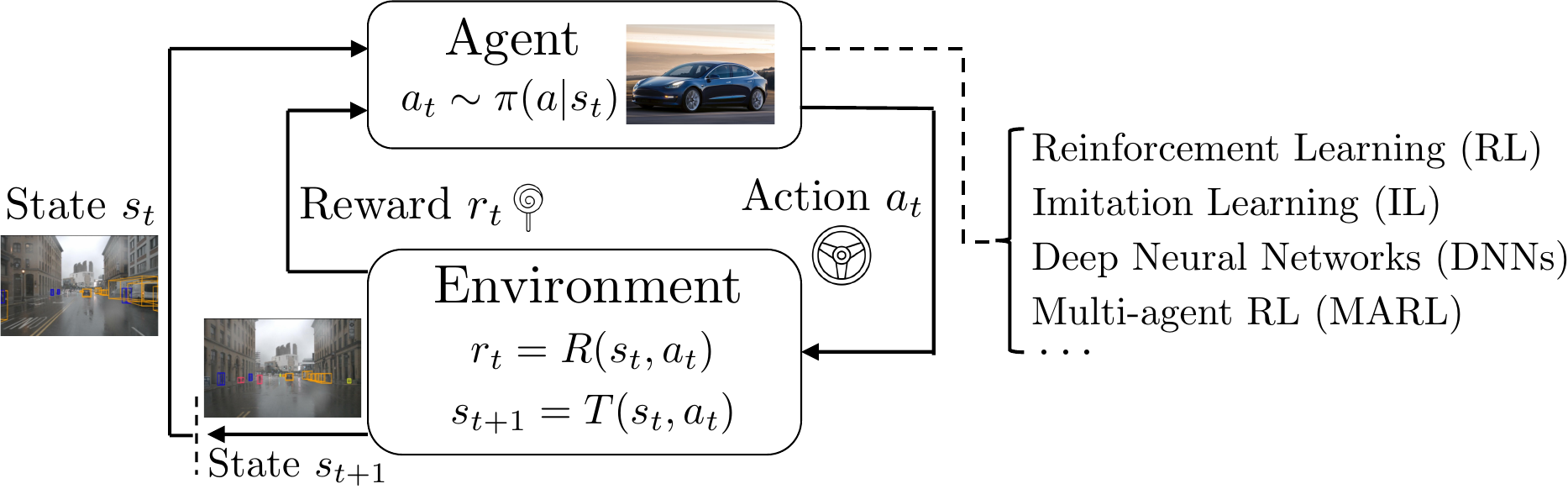}
    \caption{Agent-environment interaction in MDPs.}
    \label{fig:mdp}
\end{figure}

\noindent \textbf{Markov Decision Process (MDP).}~An MDP is a discrete-time
stochastic control process used to model sequential decision-making
problems~\cite{mdpwiki}, which comprises states $S$, actions $A$, rewards $R$,
policy $\pi$, and transitions $T$. It is thus represented as a tuple $<S, A, T,
R, \pi>$. A decision-making agent interacts with the environment at each
timestep, $t = 0, 1, 2, \cdots$, as shown in \F~\ref{fig:mdp}. At timestep $t$, the environment is in some state $s_t
\in S$, and the agent chooses an action $a_t \in A$ based on its policy $\pi$.
When the action is taken, $T$ transfers the environment to the next state. 
The agent receives an immediate reward computed by $R$ from the environment.  

\underline{States $S$.}~$S$ is a set of states, and it's also called the state
space. It can be discrete or continuous. The state $s_t \in S$ describes the
observations of the decision-making agent at timestep $t$.

\underline{Actions $A$.}~$A$ is a set of actions that an agent can take, also called
the action space, which can be discrete or continuous. 

\underline{Transitions $T$.}~$T$ is the state transition function $s_{t+1} =
T(s_t, a_t)$. The state of the MDP is $s_t$ at timestep $t$, and the agent takes
action $a_t$. The MDP will step into the next state $s_{t+1}$ according to $T$.

\underline{Rewards $R$.}~$R$ defines the immediate reward, which is also known
as the ``reinforcement.'' At state $s_t$, the agent receives immediate reward $r_t = R(s_t, a_t)$ by taking action $a_t$.

\underline{Policy $\pi$.}~$\pi$ is the agent's policy. $\pi(a|s_t)$ is the
probability distribution over possible actions. It estimates the cumulative
rewards of taking action $a_t$ at state $s_t$. A deterministic agent will take
the action that maximizes the estimated rewards according to its policy, i.e., the action with 
the highest probability given by $\pi(a|s_t)$. Meanwhile, a stochastic
agent will take an action according to the actions' probability distribution over all actions 
given by $\pi(a|s_t)$.

\underline{Markov property.}~We present the formal definition of Markov property
in Definition~\ref{def:markov}. Markov property forms the basis of an important
and unique optimization opportunity taken by \tool, as discussed in
\S~\ref{subsec:design-em}. The Markov property shows that the probability ($Pr$)
of moving to the next state $s_{t + 1}$ in an MDP depends solely on the present
state $s_{t}$ and not on the previous states.

\vspace{-3pt}
\begin{definition}[Markov property]
The sequence of the states in MDP is a Markov chain,
which has the following Markov property:
\begin{equation*}
    Pr(S_{t + 1} = s_{t + 1} | S_{t} = s_t, \cdots , S_{0} = s_0) = Pr(S_{t+1} = s_{t + 1} | S_{t} = s_{t})
\end{equation*}

\label{def:markov}
\end{definition}

\begin{figure*}[!t]
  \centering
  \includegraphics[width=1.00\linewidth]{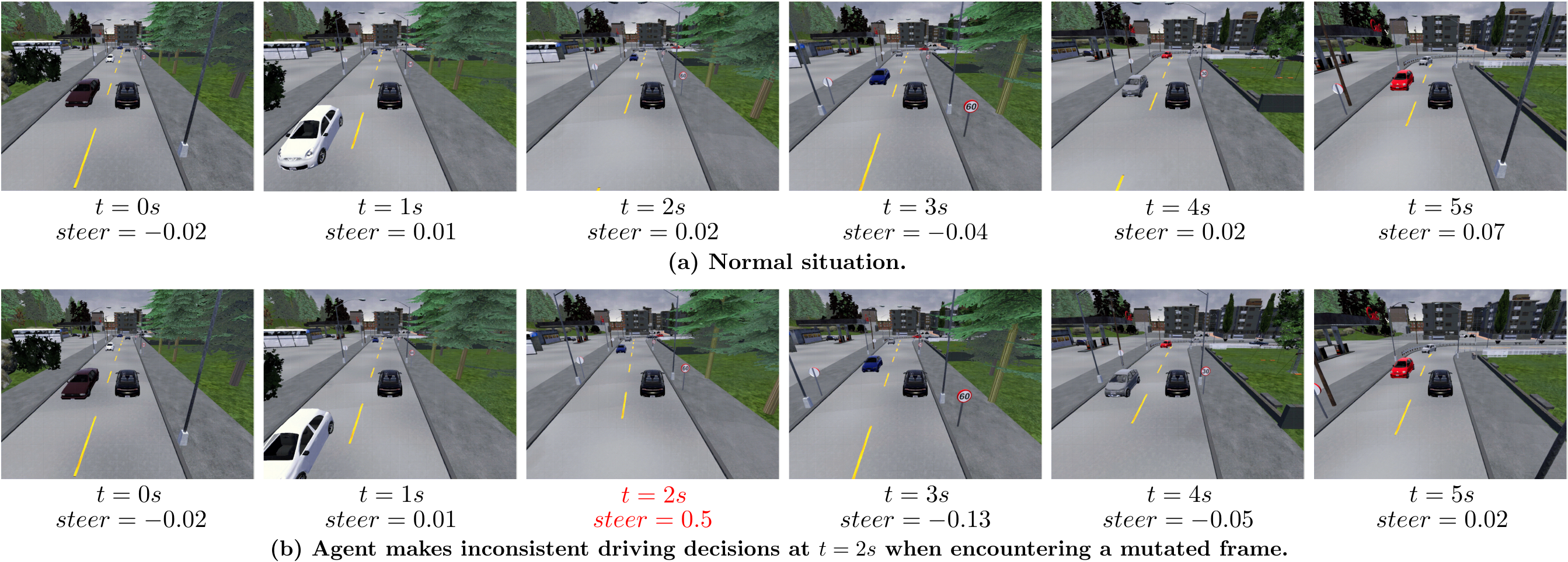}
  \caption{Inconsistencies focused by prior DNN testing do not always result
    in severe states in models solving MDPs.}
  \label{fig:example-AE}
\end{figure*}

\noindent \textbf{Models Solving MDPs.}~To date, neural models are often used to
form the agent policy $\pi$ in \F~\ref{fig:mdp}. They can thus solve sequential
decision-making problems by modeling them as MDPs and obtaining the
(nearly-)optimal policies. We now review four cutting-edge models, whose performance is 
close to or even better than humans.

\underline{Deep Neural Network (DNN).}~The supervised DNNs train an agent model
to predict the best action $a_t$ at the present state $s_t$. It assumes that the
present action $a_t$ is solely determined by the current state $s_t$, and states
before $s_t$ have no influence. With enough manually labeled data, DNNs can model an agent policy even close to the optimal policy $\pi^{*}$.
This technique has been used in autonomous driving systems developed by
NVIDIA~\cite{bojarski2016end,bojarski2017explaining} and DNN-based variant of ACAS
Xu~\cite{julian2019deep} with impressive results.

\underline{Reinforcement Learning (RL).}~Supervised DNNs often require a
significant amount of labeled data. Data labeling needs considerable human
effort, which is unrealistic in many real-world situations. RL does not
require labeled data. Instead, it uses reward functions in MDPs to guide the
agent model in estimating the cumulative reward.
A3C~\cite{mnih2016asynchronous}, DDPG~\cite{lillicrap2015continuous},
DQN~\cite{mnih2013playing}, PPO~\cite{schulman2017proximal},
TQC~\cite{kuznetsov2020controlling}, and other RL algorithms have emerged with
impressive performance. These algorithms have been deployed in complex scenarios
like Go~\cite{silver2016mastering}, video games~\cite{berner2019dota}, and robot
control~\cite{kober2013reinforcement}, and their performance has been superior
to that of humans.

\underline{Imitation Learning (IL)~\cite{ho2016generative}.}~RL does not require
a substantial amount of labeled data. However, defining reward functions
might be tricky in some cases. IL comes in handy when it is easier for an
expert to demonstrate the desired behavior than designing an explicit
reward function. The MDP is the main component of IL, and the reward function
$R$ is \textit{unknown} to the agent. The IL agent can either learn the expert's
policy or estimate the reward function by monitoring the expert's
trajectories, which are sequences of states and actions $\tau = (s_0, a_0, s_1,
a_1, \cdots)$. Waymo uses IL to learn an urban driving policy from human
drivers~\cite{bansal2018chauffeurnet}.

\underline{Multi-agent Reinforcement Learning
  (MARL)~\cite{busoniu2008comprehensive}.}~MARL provides solutions for
scenarios with multiple agents. In most cases, both the state and the reward
received by each agent are influenced by the joint actions of all agents. Agents
can cooperate to solve a problem, 
with the overall goal
being to maximize the average cumulative rewards of all agents.
The global cooperative optimum is a Nash equilibrium~\cite{osborne1994course}.
The agents can also compete with one another, resulting in a zero-sum Markov
game. MARL has been used to train agents in video
games~\cite{vinyals2019grandmaster} and traffic
control~\cite{wiering2000multi,ijcai2019-635}.

\section{Related Work}
\label{sec:related}

Existing works have laid a solid foundation in testing
DNNs~\cite{zhang2020machine}. We review these works holistically from three
aspects to present a self-contained paper. We also particularly discuss existing
works testing RL to motivate the design of \tool.

\noindent \textbf{Target DNNs.}~Recurrent neural networks (RNNs) and feedforward
neural networks (FNNs) are two representative types of DNNs. Given the wide
adoption of FNNs in computer vision (CV), most existing works test FNNs have
examined the accuracy of FNN-based image classifiers and their applications,
such as autonomous driving systems~\cite{pei2017deepxplore,zhang2018deeproad,
tian2018deeptest, dwarakanath2018identifying, nakajima2019generating,
wang2020metamorphic,yuan2021perception,yuan2021you,yuan2021enhancing}. Natural
language processing (NLP) models have also been
tested~\cite{he2020structure,sun2020automatic,ma2020metamorphic,ma2022mtteql}.
We also notice recent works on testing RNNs and RL
models~\cite{du2018deepcruiser, du2019deepstellar, guo2019rnn, huang2019testrnn,
uesato2018rigorous}. 
DeepStellar~\cite{du2019deepstellar}, a SOTA RNN testing work, models the target
RNN's internal state transition using Finite State
Transducer~\cite{gill1962introduction} as a testing guide. DeepStellar requires
a bounded RNN state space with reasonable size. This assumption may not hold for
models solving MDPs, because the state space can often be unlimited, especially
in real-world blackbox scenarios such as autonomous driving and robot control.
Moreover, the transition functions $T$ in complex real-world MDP environments
are difficult, if possible, to obtain. In contrast, \tool\ tests FNNs, RL, IL,
and MARL models for solving MDPs using a unified approach.

\noindent \textbf{Testing Oracle and Testing Criteria.}~Constructing proper
oracles has long been difficult for testing DNNs~\cite{zhang2020machine}.
Metamorphic or differential testing has been used extensively to overcome the
difficulty of explicitly establishing testing
oracles~\cite{segura2016survey,pei2017deepxplore,wang2020metamorphic,chen2021validation}.
Consequently, DNNs are considered incorrect if they produce \textit{inconsistent}
results. However, \S~\ref{sec:motivation} shows that ``inconsistency'' does not
always lead to model anomalies. For instance, an autonomous driving model can
easily recover from steering driftings. Regarding testing
criteria selection, whitebox DNN testing relies on a wide range of coverage
criteria~\cite{pei2017deepxplore,ma2018deepgauge,odena2018tensorfuzz,sun2018testing}.
In contrast, blackbox testing may use evolutionary algorithms or other
heuristics to determine test input quality~\cite{ijcai2019-800}.
\S~\ref{sec:motivation} introduces \tool's statistics-based methods for
evaluating and prioritizing test inputs.

\noindent \textbf{Input Mutation.}~Previous works testing CV models use
pixel-level mutations~\cite{pei2017deepxplore}, weather
filters~\cite{zhang2018deeproad}, and affine
transformations~\cite{tian2018deeptest} for semantics-preserving changes on
images. For natural language text, existing works often use pre-defined
templates to generate linguistically coherent text~\cite{galhotra2017fairness,
  udeshi2018automated, ma2020metamorphic, he2020structure}. DNN models typically
process each input separately. Models solving MDPs, however, constantly respond
to \textit{a series of} states, e.g., an autonomous driving model makes
decisions about each driving scene frame captured by its camera. Changing
arbitrary frames may destroy inter-state coherence; see \tool's solution in
\S~\ref{sec:motivation}.

\noindent \textbf{Existing RL Testing.}~There exist several works that test
RL~\cite{uesato2018rigorous,lee2020adaptive,julian2020validation,ernst2019fast,koren2018adaptive,yamagata2020falsification}.
\cite{uesato2018rigorous} assumes that crash patterns in weaker RL models are
similar to those in more robust RL models. During the \textit{training phase} of
RL models, crash-triggering sequences are collected to train a classifier to
predict whether a (mutated) initial state would cause abnormal future states.
This method relies heavily on historical training data, which also limits the
method's ability to discover new and diverse crash-triggering state
sequences.~\cite{lee2020adaptive,julian2020validation,ernst2019fast,koren2018adaptive,yamagata2020falsification}
adopt similar approaches to~\cite{uesato2018rigorous} to find the
crash-triggering path, but designing the scenario-specific algorithms is hard.
Moreover, the training process needs a large amount of data simulation and its
performance highly depends on the data sampling method. These works studied a
specific model/scenario, and they do not aim to deliver a unified framework to
test models solving MDPs. They train models to predict a path reaching an unsafe
state instead of detecting numerous crashes efficiently. It is hard to adapt
these methods to uncover numerous crash-triggering sequences in MDP scenarios,
considering the high training/simulation cost. Besides, while \tool\ 
leverages the Markov property, a unique property in MDP, to largely optimize the
fuzzing efficiency (see \S~\ref{subsec:design-em}), all prior works do not
consider the Markov property.

\section{Testing DNN Models Solving MDPs}
\label{sec:motivation}

This section describes the challenges that previous DNN testing works face when
testing models solving MDPs. Accordingly, we introduce several design
considerations of \tool.

\smallskip
\noindent \textbf{Inconsistencies vs.~Crashes.}~Most DNN testing works form
testing oracles by checking prediction \textit{consistency}, as reviewed in
\S~\ref{sec:related}. However, such a testing oracle is overly strict when it
comes to testing models solving MDPs.
Consider \F~\ref{fig:example-AE}, which depicts the behavior of the SOTA RL
model~\cite{toromanoff2020end} that won Camera Only track of the CARLA
challenge~\cite{carlachallenge}. We establish an autonomous driving scenario 
where the RL model decides the steer angles and speed per frame.
\F~\hyperref[fig:example-AE]{\ref{fig:example-AE}a} shows the RL model's
reaction to six frames. Then, we create an ``inconsistent'' driving behavior, at
$t = 2s$, by compelling the RL model to change its decision from virtually
straight (0.02) to turning right (0.5). Existing
works~\cite{zhang2018deeproad,tian2018deeptest} would consider such
``inconsistent'' driving behavior erroneous. However, as seen in
\F~\hyperref[fig:example-AE]{\ref{fig:example-AE}b}, the RL-controlled vehicle
quickly returns to its \textit{normal state} in the following frames, with steer
approaching zero.

A well-trained model for solving MDPs (e.g., a robot controller) can frequently
meet and recover from short-term inconsistencies. Inconsistent predictions do
not always lead to abnormal or dangerous states. We thus present the following
argument:

\begin{tcolorbox}[size=small]
  Real-world models for solving MDPs can quickly \textit{recover} from
  ``inconsistent'' predictions. This demands the creation of oracles over concrete
  and severe \textit{abnormal states}, such as collisions in airplane control or
  autonomous driving. 
\end{tcolorbox}

\begin{table}[!t]
  \centering
  \caption{Crash feasibility when the RL model yields multiple inconsistent decisions.}
  \label{tab:AE-example}
  \resizebox{0.85\linewidth}{!}{
    \begin{tabular}{|l|c|c|c|c|c|}
      \hline
      Inconsistent decision ratio & 1\% & 5\% & 10\%  & 25\% & 50\% \\
      \hline
      Crash feasibility           & 0\% & 1.5\% & 2.3\% & 6.0\% & 11.5\% \\
      \hline
    \end{tabular}
  }
\end{table}

\smallskip
\noindent \textbf{Example.}~\T~\ref{tab:AE-example} reports the relations
between the ratio of inconsistent RL-model decisions in one autonomous driving
run and the average feasibility of collisions (i.e., ``Crash feasibility''). We
run the autonomous driving model for 1,000 runs with randomly selected initial
states. Each run stops at $t=10s$, thus generating 100 frames (10 fps). We
randomly select several frames and change the RL-controlled vehicle in those
frames by compelling their steering to either left most or right most (which
rarely occur in normal driving). We find that when the RL-controlled vehicle in
a \textit{single} frame is mutated, equivalent to the 1\% inconsistent decision
ratio of \T~\ref{tab:AE-example}, it cannot result in a collision across all
1,000 runs, and the vehicle recovers quickly from inconsistent decisions. The
crash feasibility increases slightly when many decisions are changed into an
inconsistent stage in a run. Even if half of the steering decisions are changed,
only 11.5\% of these runs can cause vehicle collisions. Mutating 50\% of
decisions induce apparently unrealistic driving behaviors. Comparatively, \tool\
uncovers 164 collisions by only mutating the initial state without breaking the
naturalness (see \S~\ref{sec:implementation}) of the entire MDP procedure, as
shown in \S~\ref{sec:evaluation}.

To ease reading, this paper refers to abnormal and dangerous states as
``crashes.'' We regard crashes incurred from a valid and solvable initial state
as \textit{bugs} of a model $\pi$ solving MDP. Formally, we have the following
definition:

\begin{definition}[Crash-triggering state sequences]
\label{def:crash}
Given model $\pi$ and a valid and solvable initial state $S_0$, we observe a
state sequence: $\{S_t\}_{t\in[M-1]}$\footnote{$[M]$ is short for $\{0, 1, \cdots, M\}$ throughout this paper.}, whose actions are made by $\pi$. $S_0$ is
regarded as crash-triggering if there exists a crash state $S_t (t>0)$. 
\end{definition}

We require that $S_0$ is \textit{solvable} for an optimal model $\pi^*$, meaning
that every state in the optimal state sequence $\{S_t^*\}_{t\in[M-1]}$ is not a
crash state. Therefore, Def.~\ref{def:crash} implies that $\pi$ is buggy, whose
bug-triggering input is $S_0$.
We, clarify that ``crash'' (i.e., testing oracle) is defined case
by case; see \textbf{Testing Oracles} in \S~\ref{sec:implementation}.

\smallskip
\noindent \textbf{Mutating Initial States vs.~Mutating Intermediate
States.}~Mutating arbitrary frames in an MDP can enlarge the testing surface of
the target model and potentially reveal more defects. Mutating an intermediate
frame, however, could break the coherence of an MDP state sequence. In
\F~\ref{fig:example-AE}, our tentative study shows that by mutating the
surrounding environment in the driving scene at frame $t$, it is possible to
influence the RL-controlled vehicle's decision. However, given such mutations
introduce broken coherence when considering frames around $t$, defects found by
mutating frame $t$ may not imply real-world anomaly behaviors of autonomous
driving vehicles. We thus decide to only mutate the initial state (e.g.,
re-arrange the environment at timestep $0$), allowing to preserve the coherence
of the entire MDP state sequence. For each MDP scenario, we validate mutated
initial states in a deliberate way to ensure they are realistic and solvable for
an optimal model $\pi^*$; see \textbf{MDP Initial State Sampling, Mutation, and
Validation} in \S~\ref{sec:implementation}.

\smallskip
\noindent \textbf{Design Overview.}~\tool\ is designed to evaluate blackbox
models (not merely RL) solving MDPs. \tool\ requires no training phase
information (in contrast to~\cite{uesato2018rigorous}). Our testing oracle
checks abnormal and dangerous states instead of models' inconsistent behaviors
(see \S~\ref{sec:implementation} for our oracles). Moreover, we only mutate the
initial state rather than arbitrary states to generate more realistic scenarios.
These new designs add to the complexity and cost of testing models solving MDPs.
Worse, earlier objective-oriented generation
methodologies~\cite{pei2017deepxplore}, which rely on loss functions to directly
synthesize corner inputs, are no longer feasible because MDP transition
functions $T$ are often unavailable and not differentiable.

\tool\ offers a comprehensive and unified solution to all of the issues above.
It incorporates several optimizations to maintain and prioritize initial states
with greater potential to cover new model behaviors. Our evaluation reveals
promising findings when testing real-world models under various MDP scenarios.

\begin{figure}[!htp]
  \centering
  \includegraphics[width=1.00\linewidth]{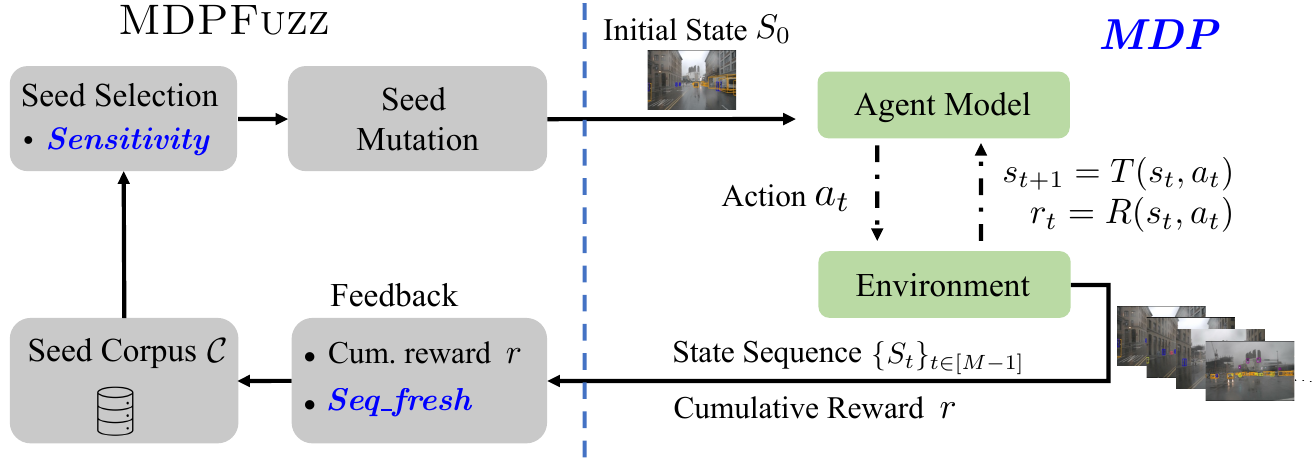}
  \caption{Workflow. The ``Agent Model'' is the testing target.}
  \label{fig:pipeline}
\end{figure}

\section{Design}
\label{sec:design}

\noindent \textbf{Assumptions.}~\F~\ref{fig:pipeline} illustrates the
workflow of \tool.
In blackbox settings, the internal of the agent model (our testing target) is
not revealed to \tool. Similarly, the MDP transition function $T$ and reward
function $R$ in the environment are unavailable. However, \tool\ can collect the
state sequence $\{S_t \}_{t \in [M-1]}$ went through by the target model and
obtain the cumulative reward $r$ from the environment.
For instance, when testing autonomous driving models, \tool\ can collect a
series of frames that the autonomous driving model captures from the environment
and obtain the cumulative reward $r$ corresponding to this series. 
We clarify that this assumption is realistic, even for (commercial) blackbox
models. As noted above, \tool\ has no access to the blackbox model internals.
However, in a typical MDP environment, the states and agent actions are
observable and the rewards are calculated based on the observed states. No
additional information is needed.
\tool\ only mutates the initial state $S_0$ of an MDP, under the constraint that
it is still realistic and solvable after mutation. See
\S~\ref{sec:implementation} on how mutated initial states are validated in this
study. 

\A~\ref{alg:fuzz-alg} formulates the fuzz testing procedure, including key
components mentioned in \F~\ref{fig:pipeline}. $\code{Fuzzing}$ is the main
entrance of \A~\ref{alg:fuzz-alg}, 
which returns the set of error-triggering initial states $\mathcal{R}$, forcing
the target model to enter severe states (e.g., collisions in autonomous
driving). As we have discussed in Definition~\ref{def:crash}, we refer to such
severe states as ``crashes'' in this paper. We summarize functions corresponding
to the key components of \F~\ref{fig:pipeline} as follows:

\begin{itemize}[leftmargin=4mm,noitemsep,topsep=0pt]
\item \code{MDP} starts from the initial state $S_0$ and observes the
  interaction between the target agent model and the environment for $M$
  timesteps (line 2). It returns the sequence of covered states and the
  cumulative reward (line 3). The length $M$ of the state sequence is a
  hyper-parameter.

\item \code{Sensitivity} estimates the sensitivity of the target model against
\tool's mutations on the initial states (see \S~\ref{subsec:design-robust}). A
larger sensitivity indicates that the model becomes less robust (a good
indicator for testing) w.r.t.~the mutated initial states. This is comparable to
estimating seed energy in software fuzzing~\cite{bohme2017coverage,7985706}.

\item \code{Seq_fresh} estimates the state sequence ``freshness'' (see
\S~\ref{subsec:design-em}). Overall, it checks whether a sequence of covered
states has new patterns that do not exist in previously-found sequences. This is
comparable to taking code coverage as feedback in software
fuzzing~\cite{afl,klees2018evaluating}. Note that \tool\ can observe state
sequence in blackbox settings.
\end{itemize}

\begin{figure}[!t]
\begin{algorithm}[H]
    \caption{\tool\ workflow}
    \footnotesize
    \label{alg:fuzz-alg}
      \begin{algorithmic}[1]
        \Function{\code{MDP}}{$S_0$}
        \LeftComment{~~$\textsc{ObserveFromEnv}$: run agent-env interaction for $M$ steps}
        \State $r$, $\{ S_t \}_{t \in [M-1]}$ $\leftarrow \textsc{ObserveFromEnv}(S_0)$
            \State \Return $r$, $\{ S_t \}_{t \in [M-1]}$
        \EndFunction

        \Function{\code{Fuzzing}}{ }
        \LeftComment{~~$\tau$: state sequence freshness threshold}
        \State $\mathcal{C} \leftarrow \textsc{Sampling}(N)$ \algorithmiccomment{Sample seed corpus with $N$ initial states}
        \State $Params^s, Params^c \leftarrow \textsc{Init()}$ \algorithmiccomment{Initialize key parameters}
        \For{$S_0^i \in \mathcal{C}$}
            \State $E_i \leftarrow \code{Sensitivity}(S_0^i)$
            \State $r_i, \{ S_t^i \}_{t \in [M-1]} \leftarrow \code{MDP}(S_0^i)$
            \State $p_i \leftarrow \code{Seq_fresh}(\{ S_t^i \}_{t \in [M-1]}, Params^s, Params^c, \tau)$
        \EndFor

        \While{\text{passed time }$< 12$ \text{hours}}
            \State Select $S_0^k$ from $\mathcal{C}$ with probability $E_k / \sum_{i=1}^N E_i$
            \State $S_{0}^{\Delta, k} \leftarrow \textsc{Mutate_Validate}(S_0^k)$
            \State $r_k^{\Delta}, \{S_{i}^{\Delta, k}\}_{i \in [M-1]} \leftarrow \code{MDP}(S_{0}^{\Delta, k})$
            \State $p_k^{\Delta} \leftarrow \code{Seq_fresh}(\{S_{i}^{\Delta, k}\}_{i \in [M-1]}, Params^s, Params^c, \tau)$
            \If{$Crash(\{S_{i}^{\Delta, k}\}_{i \in [M-1]})$} \algorithmiccomment{Testing oracles. See \S~\ref{sec:implementation} for details.}
                \State Add $S_{0}^{\Delta, k}$ to $\mathcal{R}$
            \ElsIf{$r_k^{\Delta} < r_k$ \textbf{or} $p_k^{\Delta} < \tau$} \algorithmiccomment{Feedback. See \S~\ref{subsec:design-reward} for details.}
                \State Add $S_{0}^{\Delta, k}$ to $\mathcal{C}$
                \State $E_k^{\Delta} \leftarrow \code{Sensitivity}(S_{0}^{\Delta, k})$
                \State Maintain $r_k^{\Delta}, E_k^{\Delta}, p_k^{\Delta}$ for $S_{0}^{\Delta, k}$
            \EndIf
        \EndWhile
        \State \Return $\mathcal{R}$
        \EndFunction
      \end{algorithmic}
\end{algorithm}
\setlength{\textfloatsep}{\textfloatsepsave}
\end{figure}

\noindent \textbf{Initialization.}~Lines 5--10 in \A~\ref{alg:fuzz-alg}
initializes the fuzzing campaign of \tool. Line 5 randomly samples $N$ initial
states in the legitimate state space of MDP to form the seed corpus
$\mathcal{C}$ (see \S~\ref{sec:implementation} for details). Then, the key
parameters, $Params^s$ and $Params^c$, are initialized at line 6. Holistically
speaking, these parameters maintain the up-to-date density distribution over
previously-covered sequences, and it will be used by \code{Seq_fresh} to decide
the freshness of a new sequence, as will be introduced in
\S~\ref{subsec:design-em}. We iterate each seed $S_0^i$ in corpus $\mathcal{C}$
and estimate its energy $E_i$ at line 8. Then, we feed $S_0^i$ to the target
model, receive the state sequence $\{ S_t^i \}_{t \in [M-1]}$ and its cumulative
reward $r_i$ by running \code{MDP} (line 9), and compute the sequence freshness
$p_i$ using $\code{Seq_fresh}$ (line 10).

\noindent \textbf{States Validation.}~We randomly mutate a selected seed (line
13). Seed mutation is bounded in the legitimate state space of MDP to guarantee
that such initial states exist and solvable in real-world scenarios. We also use the
validation module shipped by some MDP environments to validate the mutated
states. In short, we clarify that all the mutated initial states are
\textit{valid} and \textit{solvable}, and all the crashes we found are avoidable
if the models can take the optimal actions (though currently the tested model
failed to take the optimal actions and is thus \textit{buggy}). See
implementation details of \textsc{Mutate_Validate} in \textbf{MDP Initial State
Sampling, Mutation, and Validation} of \S~\ref{sec:implementation}.

\noindent \textbf{Fuzzing.}~As a common setup, \tool\ launches each fuzzing campaign
for 12 hours~\cite{klees2018evaluating}. Each time (line 12) we select a seed
$S_0^k$ from the corpus $\mathcal{C}$ with probability
$\frac{E_k}{\sum_{i=0}^{N-1}E_i}$, where $E_i$ denotes the energy (estimated
by \code{Sensitivity}) of the $i^{th}$ seed. Similar to software fuzzing which
spends more time on seeds of higher energy~\cite{afl,bohme2017coverage}, \tool\
prioritizes seeds with higher energy.
We feed $S_0^{\Delta, k}$ to \code{MDP}, and collect the cumulative reward
$r_k^{\Delta}$ and the state sequence $\{S_{i}^{\Delta, k}\}_{i \in [M-1]}$
(line 14). If the new initial state can cause a crash according to our oracle,
we add it to the crash-triggering set $\mathcal{R}$ (lines 16--17).

Line 15 measures the freshness of the covered state sequence, quantifying how
much the new sequence is distinct with previously-covered sequences. Then, we
check whether the new cumulative reward $r_k^{\Delta}$ is smaller than the
reward collected when using $S_0^k$, or whether the freshness is above a
threshold (line 18).\footnote{``Freshness'' is assessed via probability density.
Density $p_k^{\Delta}$ lower than $\tau$ (line 18 in \A~\ref{alg:fuzz-alg})
denotes freshness higher than a threshold; see \S~\ref{subsec:design-em} for the
details.} If so, we keep $S_0^{\Delta, k}$ in the seed corpus (line 19) and also
maintain its associated reward, energy, and sequence freshness for future usage.

\subsection{Robustness \& Sensitivity}
\label{subsec:design-robust}

\code{Sensitivity} estimates a seed's potential to provoke diverse behaviors of
the target model. Notably, the resilience of the model solving MDPs is
commonly defined in terms of their sensitivity to state permutations. Many
previous works have launched adversarial attacks by adding small permutations to
the observed states of RL
models~\cite{xiao2019characterizing,pattanaik2017robust,
behzadan2017vulnerability, kos2017delving}.

\begin{figure}[!t]
\setlength{\textfloatsep}{0pt}
\begin{algorithm}[H]
    \caption{Sensitivity estimation.}
    \footnotesize
    \label{alg:fuzz-uncertainty}
      \begin{algorithmic}[1]
        \Function{\code{Sensitivity}}{$S_0$}
            \State $S_0^{\Delta} \leftarrow S_0 + \Delta S$ \algorithmiccomment{$\Delta S$ is a small random permutation}
            \State $r, \{ S_t \}_{t \in [M-1]} \leftarrow \code{MDP}(S_0)$,
            \; $ r_{\Delta}, \{ S_t^{\Delta} \}_{t \in [M-1]} \leftarrow \code{MDP}(S_0^{\Delta})$
            \State \Return $\frac{|r - r_{\Delta}|}{||\Delta S||_2}$
        \EndFunction
      \end{algorithmic}
\end{algorithm}
\setlength{\textfloatsep}{\textfloatsepsave}
\end{figure}

Inspired by these works, the potential of a seed, i.e., an initial state $S_0$,
is estimated by the sensitivity of the target model w.r.t. randomness in $S_0$. As shown in
\A~\ref{alg:fuzz-uncertainty}, \code{Sensitivity} adds small random permutations
$\Delta S$ to an initial state $S_0$ and then collects the cumulative reward
$r_{\Delta}$ from \code{MDP}. The local sensitivity of the model at $S_0$ can
thus be estimated by $\frac{|r - r_{\Delta}|}{||\Delta S||_2}$, where $r$ is the
cumulative reward without permutation. 
Interested readers may refer to an illustrative example of \A~\ref{alg:fuzz-uncertainty}'s intuition at~\cite{snapshot}.

\subsection{State Sequence Freshness \& DynEM}
\label{subsec:design-em}

\code{Seq_fresh} in \A~\ref{alg:fuzz-alg} guides \tool\ to promptly identify
``fresh'' state sequences. Software fuzzing keeps mutated seeds if new coverage
patterns are exposed~\cite{afl}. Similarly, \tool\ keeps a mutated initial state
if its induced state sequence is distinct from previously covered sequences
(line 18 in \A~\ref{alg:fuzz-alg}). Neuron coverage~\cite{pei2017deepxplore}, a
common criterion in DNN testing, is not proper here; see \S~\ref{sec:discussion}.

\noindent \textbf{Motivation.}~A naive way to measure the freshness between a
new state sequence and previously covered sequences is to calculate the minimum
distance between them iteratively. \T~\ref{tab:EMtime} compares a naive
distance-based method with \code{Seq_fresh}. We use the setting of RL for CARLA
assessed in \S~\ref{sec:motivation}, where the state sequence length is 100, and
each agent state's dimension is 17. Given a new state sequence, the
distance-based method iterates all historical sequences (``corpus size'' in
\T~\ref{tab:EMtime}) for comparison. Calculating the distance between a new
sequence and all existing sequences in the corpus takes roughly a minute when
the corpus size is 1,000, which is too costly considering that there are
thousands of mutations in a 12-hour run, as will be reported in
\S~\ref{subsec:evaluation-RQ1}.

\noindent \textbf{Estimating Freshness with Constant Cost.}~\tool\ proposes to
first estimate a probability density function (pdf) with existing state
sequences. Then, comparing a new state sequence with existing sequences is
recast to calculating the density of the new sequence by the pdf. A new sequence
emitting low density indicates that it's not covered by existing sequences. That
is, \textit{low density indicates high freshness}.

Enabled by DynEM (introduced soon), \tool\ can measure the freshness of a new
sequence with the pdf of all existing sequences in \textit{one run}. Further,
benefiting from the Markov property introduced in Definition~\ref{def:markov},
we only need to estimate two separate pdfs with much smaller input spaces than
the entire sequence. The cost of each comparison becomes thus \textit{constant}
and \textit{modest}.
In reality, \code{Seq_fresh} only takes 0.25 seconds for tasks benchmarked in
\T~\ref{tab:EMtime} and does not scale with corpus size. Recent software
fuzzing~\cite{manes2020ankou} uses PCA-based approaches~\cite{wold1987principal}
which are not applicable here; see discussion in \S~\ref{sec:discussion}.

\noindent \textbf{\code{Seq_fresh}.}~Overall, \A~\ref{alg:fuzz-dynamicEM}
estimates $Pr(S_t)$ and $Pr(S_{t+1}, S_{t})$, as shown in the following joint
pdf of a state sequence:

\scriptsize
\begin{equation}
    \label{eq:pdf}
    \begin{split}
        Pr(\{ S_t \}_{t \in [M]})
         =  Pr(S_{M}, \cdots , S_{0})
        = Pr(S_{0}) \prod_{t = 0}^{M-1} Pr(S_{t+1} | S_{t})
        = Pr(S_{0}) \prod_{t = 0}^{M-1} \frac{Pr(S_{t+1}, S_{t})}{Pr(S_{t})}
\end{split}
\end{equation}
\normalsize

Aligned with convention~\cite{belghazi2018mutual, fei2006one, hjelm2018learning}
in machine learning, we use Gaussian mixture models
(GMMs)~\cite{mclachlan1988mixture} to estimate the pdfs, given that GMMs can
estimate any smooth density distribution~\cite{goodfellow2016deep}. Then,
\E~\ref{eq:pdf} is re-written in the following form:

\scriptsize
\begin{equation}
\label{eq:pdfgmm}
\begin{split}
    Pr&(\{ S_t \}_{t \in [M]}) = GMM^s(S_0) \prod_{t=0}^{M-1}\frac{GMM^c(S_{t+1}||S_t)}{GMM^s(S_t)} \\
    &= (\sum_{k=0}^{\mathcal{K}-1} \phi_k^s \mathcal{N}(S_0 \, | \, \mu_k^s, \Sigma_k^s)) 
    \prod_{t=0}^{M-1} \frac{\sum_{k=0}^{\mathcal{K}-1} \phi_k^c \mathcal{N} (S_{t+1}||S_{t} \, | \, \mu_k^c, \Sigma_k^c)}{\sum_{k=0}^{\mathcal{K}-1} \phi_k^s \mathcal{N}(S_t \, | \, \mu_k^s, \Sigma_k^s)}
\end{split}
\end{equation}
\normalsize

\noindent, where $\mathcal{N}$ is the pdf of Gaussian distribution, $\{\phi_k^s, \mu_k^s, \Sigma_k^s\}_{k \in [\mathcal{K} - 1]}$
are parameters of the GMM estimating single state pdf $Pr(S_t)$, $s$ in the
superscript denotes ``single,'' $\{\phi_k^c, \mu_k^c, \Sigma_k^c\}_{k \in
  [\mathcal{K} - 1]}$ are parameters of the GMM estimating concatenated states
pdf $Pr(S_{t+1}, S_t)$, and $c$ denotes ``concatenated.''
\E~\ref{eq:pdfgmm} forms the basis of \code{Seq_fresh} as in
\A~\ref{alg:fuzz-dynamicEM}. Thus, to calculate \code{Seq_fresh}, we only need
the parameters of the two GMMs: $\{\phi_k^s, \mu_k^s, \Sigma_k^s, \phi_k^c,
\mu_k^c, \Sigma_k^c \}_{k \in [\mathcal{K} - 1]}$ in \E~\ref{eq:pdfgmm}. We
develop \textsc{DynEM} to estimate and update these parameters efficiently based
on the state sequences \tool\ has already covered.

\begin{table}[!t]
  \centering
  \caption{Cost of distance-based method and \code{Seq_fresh}.}
  \label{tab:EMtime}
  \resizebox{1.00\linewidth}{!}{
    \begin{tabular}{|l|c|c|c|c|}
      \hline
      \textbf{Corpus size} & 100 & 500 & 1,000 & 3,000 \\
      \hline
      \textbf{Processing time (sec) of distance-based method} & 5.31 & 26.10 & 54.19 & 156.87  \\
      \hline
      \textbf{Processing time (sec) of \code{Seq_fresh}} & \multicolumn{4}{|c|}{0.25} \\
      \hline
    \end{tabular}
  }
\end{table}

\begin{figure}[!t]
\setlength{\textfloatsep}{0pt}
\begin{algorithm}[H]
  \caption{State sequence freshness.}
  \footnotesize
  \label{alg:fuzz-dynamicEM}
    \begin{algorithmic}[1]
      \Function{\code{Seq_fresh}}{$\{S_t\}_{t \in [M-1]}$, $Params^s$, $Params^c$, $\tau$}

          \For{$t \in [M-2]$}
              \State $p(S_t) \leftarrow \textsc{GMM}(S_t, Params^s)$
              \State $p(S_{t}, S_{t+1}) \leftarrow \textsc{GMM}(S_{t} || S_{t+1}, Params^c)$
          \EndFor
          \State $p \leftarrow p(S_0) \times \prod_{t = 0} ^ {M - 2} \frac{p(S_{t}, S_{t+1})}{p(S_{t})}$
          \If{$p < \tau$}
              \State $Params^s, Params^c \leftarrow \textsc{DynEM}(\{S_t\}_{t \in [M-1]}, Params^s, Params^c)$
          \EndIf
          \State \Return $p$
      \EndFunction

    \end{algorithmic}
\end{algorithm}
\setlength{\textfloatsep}{\textfloatsepsave}
\end{figure}

\noindent \textbf{\textsc{DynEM}.}~Unlike expectation-maximization
(EM)~\cite{dempster1977maximum}, which re-computes the estimation whenever the
seed corpus is updated, \textsc{DynEM} in \A~\ref{alg:fuzz-dynamicEM} estimates
the GMM parameters in an online and incremental manner. Thus, \textsc{DynEM}
greatly reduces the computation cost of EM while achieving the asymptotic
equivalence to the EM algorithm~\cite{cappe2009line}. We adopt the ideas
proposed in the online EM algorithm~\cite{cappe2009line} and extend it to
estimate the parameters of GMMs. Instead of calculating the parameters of GMMs
directly, \textsc{DynEM} updates the complete and sufficient (C-S)
statistics~\cite{casella2021statistical} every time \tool\ finds a new state sequence. 
In short, to estimate the parameters of the pdf, we only need its corresponding
C-S statistics, and the parameters of \E~\ref{eq:pdfgmm} can be directly derived
from the C-S statistics parameters of \textsc{DynEM}, which contain all the information we need and correspond to
$Params^s$ and $Params^c$.
Also, when \tool\ finds a new state sequence, $Params^s$ and $Params^c$ are updated in \textsc{DynEM}, and consequently,
updating the parameters of \E~\ref{eq:pdfgmm}. This way, \code{Seq_fresh} can
maintain the up-to-date pdf of all covered sequences.
Details of \textsc{GMM} and \textsc{DynEM} are omitted; see the full algorithm,
the  explanation of C-S statistics, and clarification on all the employed
statistics mechanisms at our artifact~\cite{snapshot}.

$Params^s$ and $Params^c$ in \A~\ref{alg:fuzz-dynamicEM} are randomly
initialized (line 6 in \A~\ref{alg:fuzz-alg}). When a new state sequence $\{
S_{t} \}_{t \in [M]}$ is found by \tool, its sequence density $p$ is calculated
following \E~\ref{eq:pdfgmm} (lines 3--5 in \A~\ref{alg:fuzz-dynamicEM}).
If its density $p$ is smaller than the threshold $\tau$, meaning that we find a
fresh sequence, we use \textsc{DynEM} to update the key parameters, thereby
updating our maintained pdf of existing state sequences (lines 6--7 in
\A~\ref{alg:fuzz-dynamicEM}).

\subsection{Feedback from Freshness \& Reward}
\label{subsec:design-reward}

The fuzzing feedback is derived from the state sequence freshness and the
cumulative reward. This corresponds to line 18 in \A~\ref{alg:fuzz-alg}.

\begin{figure}[!t]
    \centering
    \includegraphics[width=0.98\linewidth]{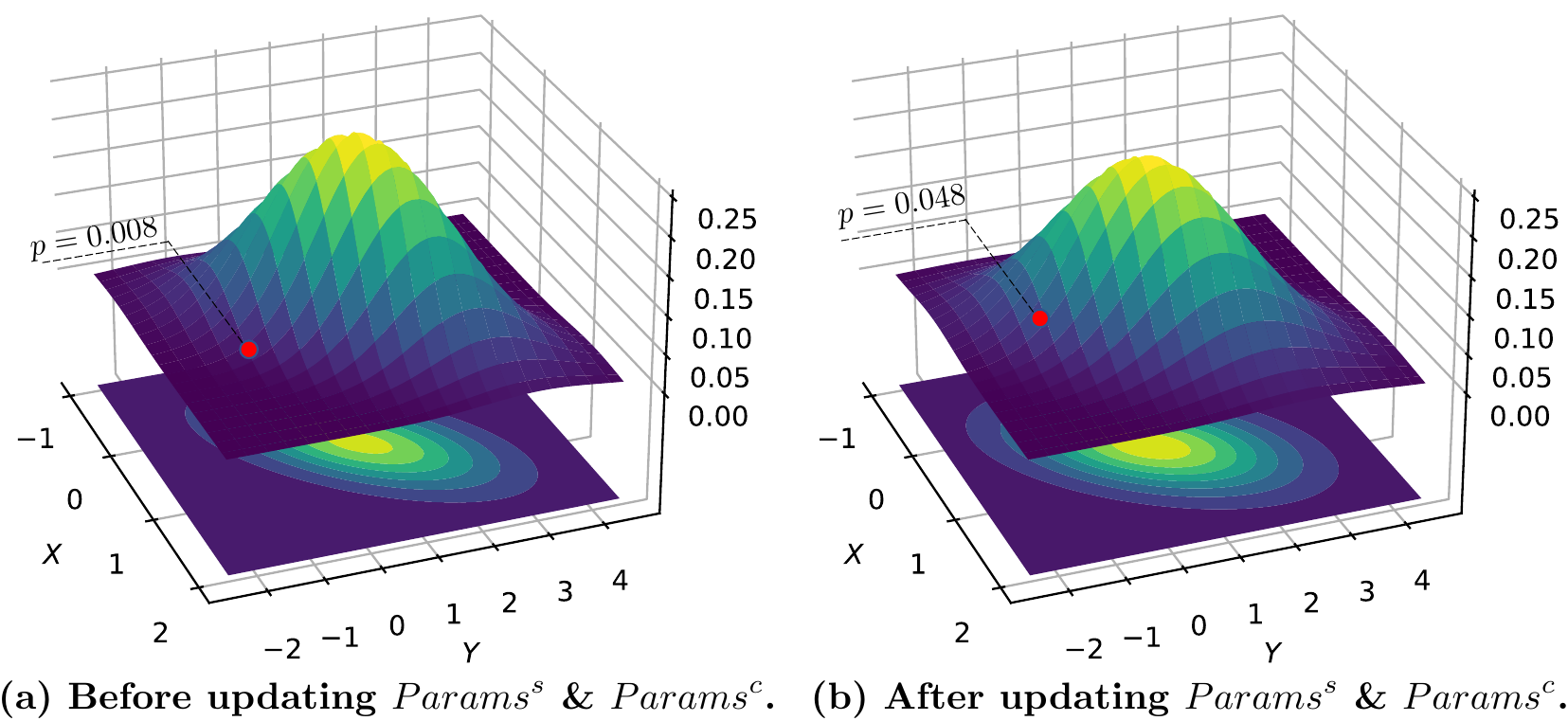}
    \caption{Illustration of \A~\ref{alg:fuzz-dynamicEM}.}
    \label{fig:EM-fig}
\end{figure}

\noindent \textbf{Freshness.}~\F~\ref{fig:EM-fig} illustrates the use of
freshness to guide fuzzing. When \tool\ finds a new state sequence (the
\textcolor{red}{red dot} in \F~\ref{fig:EM-fig}), we compute the sequence
density ($p=0.008$; line 15 in \A~\ref{alg:fuzz-alg}), which is lower than the
threshold $\tau=0.01$. The mutated initial state is then added to the corpus
(line 19 in \A~\ref{alg:fuzz-alg}), and \textsc{DynEM} also updates 
$Params^s$ and $Params^c$.
Following this update (\F~\hyperref[fig:EM-fig]{\ref{fig:EM-fig}b}), the
newly-discovered sequence density increases to $p=0.048$. In the future, similar
sequences will have densities of approximately $p=0.048$ (greater than $\tau$),
and they will not be added to the corpus unless they can reduce the cumulative
reward (see below). When \textsc{DynEM} is constantly fed with new sequences,
the estimated sequence density distribution becomes increasingly wider, e.g.,
comparing the covered area by the distribution in
\F~\hyperref[fig:EM-fig]{\ref{fig:EM-fig}b} with that in
\F~\hyperref[fig:EM-fig]{\ref{fig:EM-fig}a}. This reflects that \tool\ keeps
finding new state sequences that can occur in MDP. The efficiency of \code{Seq_fresh} is evaluated in \S~\ref{subsec:evaluation-RQ2}.

\noindent \textbf{Cumulative Reward.}~\tool\ is also guided by cumulative
rewards. Typically, when a
model is trained for solving MDPs, reward functions are needed to \textit{maximize} the state
sequences' cumulative rewards. For example, autonomous driving models are 
penalized, with small (negative) rewards, if they collide, run red lights, violate the speed limit, or commit other infractions. Models are rewarded
positively if they follow the scheduled routine and move towards the
destination.
We use the cumulative reward to quantify the target model's behavior. A low
cumulative reward suggests a high risk of catastrophic failures.
\tool\ prioritizes mutated initial states that can reduce the cumulative
rewards. Particularly, let an initial state $S_0^{\Delta, k}$ be mutated from
$S_0^k$, we retain $S_0^{\Delta, k}$ in the seed corpus if the new state sequence
starting from $S_0^{\Delta, k}$ induces a lower cumulative reward than that of
$S_0^k$ (line 18 in \A~\ref{alg:fuzz-alg}).

\section{Implementation \& Evaluation Setup}
\label{sec:implementation}

\begin{figure*}[!t]
  \centering
  \includegraphics[width=1.00\linewidth]{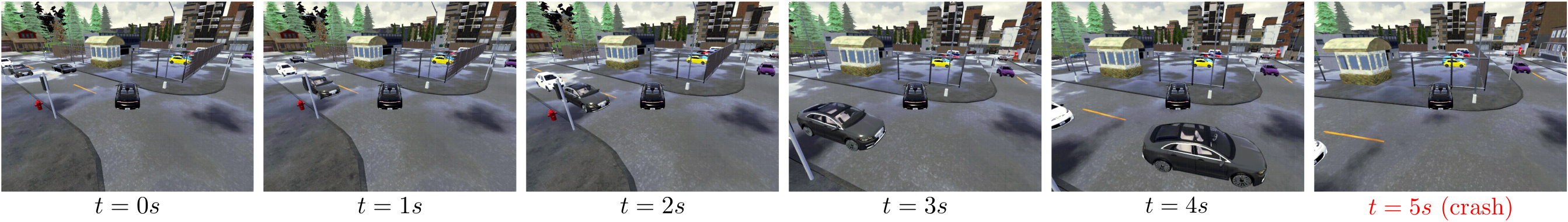}
  \vspace{-10pt}
  \caption{Crash triggering sequence found by \tool. (Please refer to~\cite{snapshot} for more results.)}
  \label{fig:crash-case1}
  \vspace*{-6pt}
\end{figure*}

\tool\ is written in Python with approximately 1K LOC. It can be integrated to
test different models solving MDPs. We use Scipy (ver.~1.6) and NumPy
(ver.~1.19) for GMM and \textsc{DynEM} calculation. We run all tested models on
PyTorch (ver.~1.8.0). Given an initial state $S_0$, \code{MDP} in
\A~\ref{alg:fuzz-alg} measures $M$ sequential states, and we set adequately
large $M$ for different models. In short, if the agent's states change rapidly,
we may use a smaller $M$, and vice versa. \T~\ref{tab:results} reports $M$
(``\#Frames'') for each setting. Users can increase $M$ to stress particularly
robust models. Each fuzzing campaign takes 12 hours, as a standard
setting~\cite{klees2018evaluating}. Before that, we randomly sample for two
hours to create the initial seed corpus (lines 5--10 in \A~\ref{alg:fuzz-alg}).
We report that our sampled initial seeds produce almost no crashes for
our tested models, e.g., throughout the 2 hours of seed sampling of CARLA RL, 
just one state sequence triggered a crash, and similar observations were made over other datasets.
Experiments are launched on a machine with one AMD Ryzen
CPU, 256GB RAM, and one Nvidia GeForce RTX 3090 GPU.

\noindent \textbf{Target Models and Environments.}~CARLA~\cite{dosovitskiy2017carla} is a popular autonomous
driving simulator. \tool\ tests SOTA RL and IL 
models~\cite{toromanoff2020end, chen2020learning} in CARLA. The RL 
model won the Camera Only track of the CARLA competition~\cite{carlachallenge},
and the IL model is currently ranked \#1 in the CARLA
leaderboard~\cite{carlaleadboard}. They determine steering and acceleration
using images collected by the vehicle's camera as inputs. ACAS
Xu~\cite{marston2015acas} is a collision avoidance system for airplanes.
In this work, we focus on the DNN-based variant of ACAS Xu~\cite{julian2019deep}, which has promising performance and much less memory requirement than the original ACAS Xu. 
It is also well-studied by existing DNN verification work~\cite{wang2018formal}.
The DNN-based ACAS Xu uses 45 distinct neural networks to predict the best actions, such as
\textit{clear of conflict}, \textit{weak/strong left and right turns}. 
Cooperative
Navigation (Coop Navi)~\cite{lowe2017multi} is an OpenAI-created  environment for
MARL. Coop Navi requires agents to cooperate to reach a set of
landmarks without colliding. We use OpenAI's release code to train the model to
the performance stated in their article~\cite{lowe2017multi}. The MARL models
use each agent's position and target landmarks to decide their actions (e.g.,
moving direction and speed). Another RL model is for the OpenAI Gym BipedalWalker environment~\cite{kuznetsov2020controlling}. In BipedalWalker, the agent
attempts to walk through grasslands, steps, pits, and stumps. We use the
publicly available TQC~\cite{kuznetsov2020controlling} model from the well-known
stablebaseline3 repository~\cite{rl-zoo3}, which takes a 24-dimension state as
input and predicts the speed for each leg based on body angle, leg angles,
speed, and lidar data.

\noindent \textbf{Testing Oracles (Crash Definition).}~For RL and IL models in
CARLA, we examine whether the model-controlled vehicle collides with other
vehicles or buildings. 
For DNN-based ACAS Xu, we check collisions of the DNN-controlled airplane with
other airplanes. For Coop Navi, we check collisions between MARL-controlled
agents. For BipedalWalker, we check whether the RL-controlled walking agent
falls (i.e., its head touches the ground). 

As expected, the crash or abnormal state is defined case by case, and orthogonal
to \tool. Users can configure \tool\ with other undesired behaviors as long as
such abnormal states can induce reasonably low rewards. In sum, we deem that the
abnormal state definition is not restricted to certain scenarios, and we
evaluate \tool\ on multiple models across multiple scenarios to alleviate
potential concern that \tool\ won't generalize. 

\noindent \textbf{MDP Initial State Sampling, Mutation, and Validation.}~As
aforementioned, the initial states of MDPs (e.g., the positions of all
participants) serve the test inputs. \tool\ samples (line 5 in
\A~\ref{alg:fuzz-alg}) and mutates (line 13) MDP initial states. In brief, we
emphasize that: 

\begin{tcolorbox}[size=small]
  When sampling, the initial states are randomly selected from the state space
  $\mathcal{P}$. \tool\ randomly mutates initial states with small noise. All
  the mutated states are validated in an automated and deliberate way (see
  below). That is, we confirm that all mutated states are valid and solvable.
\end{tcolorbox}

We add random noise to the initial states and the noise type depends on type of
the original data, e.g., adding small Gaussian float numbers from $N(0,1)$ to
vehicles' initial positions, and adding small uniform integers from $U(0,4)$ to
the ground type in BipedalWalker. We use a similar strategy to add small random
permutations in \code{Sensitivity}. 

Particularly, in CARLA, we change the initial positions and angles of all 100
vehicles, including the model-controlled vehicle. Note that CARLA validates and
rejects abnormal initial states: \textit{all} mutated initial states passed its
validation. The model-controlled car's initial speed is zero, and the
environments guarantee other vehicles won't cause crashes, such that the optimal
actions (e.g., stop) can always avoid the crash. 
In DNN-based ACAS Xu, we mutate the initial positions and speeds of the
model-controlled and the other planes. Moreover, we bound the maximal speed of
all airplanes below 1,100 ft/sec, which is within the range of normal speed that
a plane can reach in DNN-based ACAS Xu. We guarantee that there exist optimal actions to avoid the crash and solve the
initial states, and we do not use an initial state that is not solvable.
In Coop Navi, we mutate the initial positions of the three agents controlled by
MARL. These initial positions prevent agents from colliding, and their initial
speeds are 0. Our mutated initial states can pass Coop Navi's initial state
validation module, and we confirm there exist optimal solutions to avoid
the crashes for our mutated initial states.
In BipedalWalker, we modify the environment's ground by mutating the sequence of
the ground type the agent meets, e.g., in ``flat, $\cdots$, stairs, flat, stump,
$\cdots$'', the first 20 frames are ``flat'' to ensure that the agent does not
fail initially. We then place a ``flat'' between two hurdles such that the agent
can pass the obstacles when taking optimal actions.

\section{Evaluation}
\label{sec:evaluation}

\noindent \textbf{Overview.}~We have reported the evaluation setup in
\S~\ref{sec:implementation}. In evaluation, we mainly explore the following
research questions.
\textbf{RQ1}: Can \tool\ efficiently find crash-triggering state sequences from multiple
SOTA models that solve MDPs in varied scenarios? 
\textbf{RQ2}: Can \tool\ be efficiently guided using the state sequence
freshness (\code{Seq_fresh}), and can it cover more state sequences than using
cumulative reward-based guidance alone?
\textbf{RQ3}: What are the characteristics and implications of crash-triggering
states?
\textbf{RQ4}: Can we use \tool's findings to enhance the models' robustness? 
We answer each research question in one subsection below.

\begin{table}[!t]
  \centering
  \caption{Result overview}
  \vspace{-10pt}
  \label{tab:results}
  \resizebox{1.0\linewidth}{!}{
    \begin{tabular}{|c|c|c|c|c|}
      \hline
      \textbf{Model} & \textbf{MDP Scenario} & \textbf{\#Frames} & \textbf{\#Mutations} & \textbf{\#Crashes} \\
      \hline
      RL &  CARLA autonomous driving & 100 & 3,476.7 ($\pm$ 165.3) & 161.7 ($\pm$ 10.4)  \\
      \hline
      \multirow{2}{*}{DNN} & DNN-based ACAS Xu & \multirow{2}{*}{100} & \multirow{2}{*}{161,542.7 ($\pm$ 1,226.2)} & \multirow{2}{*}{135.7 ($\pm$ 12.7)} \\
                           & aircraft collision avoidance & & & \\
      \hline
      IL &  CARLA autonomous driving & 200 & 3,166.7 ($\pm$ 25.0) & 88.0 ($\pm$ 7.0)  \\
      \hline
      MARL & Coop Navi game & 100 & 531,200.3 ($\pm$ 1,983.5) & 80.7 ($\pm$ 4.9) \\
      \hline
      RL & BipedalWalker game & 300 & 6,602.0 ($\pm$ 149.5) & 124.0 ($\pm$ 10.0) \\
      \hline
    \end{tabular}
  }
  \vspace{-10pt}
\end{table}

\begin{figure}[!b]
  \vspace{-5pt}
  \centering
  \resizebox{1.0\linewidth}{!}{\pgfplotstableread[row sep=\\,col sep=&]{
    model & T1 & T2 & T3 \\
    RL for CARLA  & 12 & 11.4517  & 0.5483  \\
    DNN-ACAS Xu (Real) & 12 & 11.7612  & 0.2368  \\
    DNN-ACAS Xu (Simulate) & 12 & 2.85  & 9.15 \\
    IL for CARLA       & 12 & 11.2137 & 0.7863 \\
    MARL (Real)   & 12 & 10.7979 & 1.2021 \\
    MARL (Simulate)   & 12 & 2.1259 & 9.8741 \\
    RL for BipedalWalker   & 12  & 8.21 & 3.79 \\
    }\mydata

\begin{tikzpicture}
    \begin{axis}[
            ybar,
            bar width=0.6cm,
            width=\textwidth,
            height=.3\textwidth,
            legend style={font=\Large, at={(0.5,1.16)},
                anchor=north,legend columns=-1},
            symbolic x coords={RL for CARLA,DNN-ACAS Xu (Real),IL for CARLA,MARL (Real),RL for BipedalWalker,DNN-ACAS Xu (Simulate),MARL (Simulate)},
            xtick=data,
            xticklabel style={font=\Large,rotate=-12},
            nodes near coords,
            nodes near coords align={vertical},
            nodes near coords style={font=\Large},
            ymin=0,ymax=14,
            ylabel={\Large Time (h)},
            xticklabel style={font=\Large},
        ]
        \addplot table[x=model,y=T1]{\mydata};
        \addplot table[x=model,y=T2]{\mydata};
        \addplot table[x=model,y=T3]{\mydata};
        \legend{Total time, Agent computation and Env interaction, \tool\ computation}
    \end{axis}
\end{tikzpicture}}
  \vspace{-25pt}
  \caption{Running time of \tool.}
  \label{fig:timing}
  \vspace{-5pt}
\end{figure}

\subsection{RQ1: Performance on Finding Crashes}
\label{subsec:evaluation-RQ1}

\noindent \textbf{Setup.}~We use the evaluation setup described in
\S~\ref{sec:implementation}. That is, we launch \tool\ to fuzz each MDP model
(listed in \T~\ref{tab:results}) and detect crashes. We collect all
error-triggering inputs for analysis.

\noindent \textbf{Results.}~\T~\ref{tab:results} summarizes the findings of
fuzzing each model. 
We report the average mutation and crash numbers and the corresponding standard deviation of three runs.
Overall, \tool\ detects a significant number of crashes from
all test cases. Within the 12-hour fuzzing, \tool\ generates more mutated
initial states for the DNN-based ACAS Xu and Coop Navi cases. This is because
the agents take less time to interact with the environments in these two cases;
see the processing time evaluation below.
\F~\ref{fig:crash-case1} reports a crash found by \tool\ on the RL model used
for the CARLA autonomous driving scenario. At timestep $t=0s$ (our mutated initial state),
the speed of the agent vehicle (the black car in the middle of \F~\ref{fig:crash-case1}) is $0$, and it has no collisions with any other
vehicles or buildings. During timesteps $t=1 \sim 5s$, we observe that the RL model accelerates the vehicle without adjusting its steering. Hence, instead of turning left or moving
in reverse, the vehicle hits the fence when $t=5s$.
Overall, \tool\ can consistently reveal defects of models solving MDPs despite
that they have different model paradigms or under distinct scenarios. We view
this illustrates the strength and generalization of \tool. On the other hand,
previous works have rarely focused on or systematically tested these models.
Ignoring their potential defects will likely result in fatal incidents in
real-world autonomous driving, airplane control, and robot control systems. 
We present the videos of the crashes found in other models
at~\cite{snapshot}.

We report the time spent on each model in \F~\ref{fig:timing}.
In most cases, the target model's computation and the interaction between the
agent and the environment occupy most time. \tool\ only introduces a small
overhead, considering that interaction in complex MDP environments is very
time-consuming. Unlike other systems, the environments of DNN-based ACAS Xu and Coop Navi are
simpler and do not consider physical effects. Therefore, these two
systems can be accelerated, where one second in the real-world only requires
approximately 6.36 ms and 23.97 ms in the DNN-based ACAS Xu (Simulate) and Coop
Navi (Simulate), respectively. We accordingly set up these two simulation
systems and re-run 12-hour fuzzing, whose results are also reported in
\F~\ref{fig:timing}. As expected, more computational resources are allocated to
\tool.
In sum, we deem that the cost of \tool\ is reasonable, especially when testing models solving
complex MDPs. We also encourage users to configure their target 
environment in the ``simulation'' mode whenever possible to leave \tool\ more
time for fuzzing.

\vspace{-2pt}
\begin{tcolorbox}[size=small]
\textbf{Answer to RQ1}: \tool\ efficiently finds crash-triggering state sequences on
models solving MDPs with modest overhead.
\end{tcolorbox}
\vspace{-2pt}

\subsection{RQ2: Efficiency \& State Coverage}
\label{subsec:evaluation-RQ2}

\noindent \textbf{Setup.}~We use the same setting as that used in
\S~\ref{subsec:evaluation-RQ1}, and we compare the performance of \tool\ with
and without the guidance of state sequence freshness computed by
\code{Seq_fresh}. We run fuzz testing with these two versions of \tool\ on the
models mentioned in \S~\ref{sec:implementation} for 12 hours. As we mentioned in
\S~\ref{subsec:design-em}, the GMMs and \textsc{DynEM} are used to estimate the
state sequence freshness. Thus, we can compare the covered areas of the two
fuzzers by visualizing the distributions of their estimated GMMs.

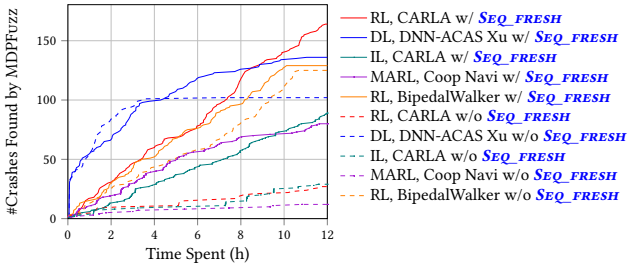
\begin{figure}[!t]
    \centering
    \resizebox{1.00\linewidth}{!}{\begin{tikzpicture}
    \begin{axis}[
      grid=major,
      height=.36\textwidth,
      xmin=0, xmax=12, ymin=0,
      legend cell align={left},
      ytick align=outside, ytick pos=left,
      xtick align=outside, xtick pos=left,
      xlabel={\Large Time Spent (h)},
      ylabel={\Large \#Crashes Found by \tool},
      legend pos=outer north east,
      legend style={draw=none}]

\addplot+[
  red, mark options={scale=0},
  error bars/.cd, 
      y fixed,
      y dir=both, 
      y explicit
  ] table [x expr=\thisrow{x}/3600, y=y] {fig/RL-carla.txt};
  \addlegendentry{\Large RL, CARLA w/ \code{Seq\_fresh}}

\addplot+[
blue, mark options={scale=0},
error bars/.cd, 
    y fixed,
    y dir=both, 
    y explicit
] table [x expr=\thisrow{x}/3600, y=y] {fig/beforerepair.txt};
\addlegendentry{\Large DL, DNN-ACAS Xu w/ \code{Seq\_fresh}}

\addplot+[
teal, mark options={scale=0},
error bars/.cd, 
    y fixed,
    y dir=both, 
    y explicit
] table [x expr=\thisrow{x}/3600, y=y] {fig/IL-carla.txt};
\addlegendentry{\Large IL, CARLA w/ \code{Seq\_fresh}}

\addplot+[
purple, mark options={scale=0},
error bars/.cd, 
    y fixed,
    y dir=both, 
    y explicit
] table [x expr=\thisrow{x}/3600, y=y] {fig/MARL.txt};
\addlegendentry{\Large MARL, Coop Navi w/ \code{Seq\_fresh}}

\addplot+[
orange, mark options={scale=0},
error bars/.cd, 
    y fixed,
    y dir=both, 
    y explicit
] table [x expr=\thisrow{x}/3600, y=y] {fig/RL-Game-EM.txt};
\addlegendentry{\Large RL, BipedalWalker w/ \code{Seq\_fresh}}

\addplot+[
dashed, red, mark options={scale=0},
error bars/.cd, 
    y fixed,
    y dir=both, 
    y explicit
] table [x expr=\thisrow{x}/3600, y=y] {fig/RL-carla-noEM.txt};
\addlegendentry{\Large RL, CARLA w/o \code{Seq\_fresh}}

\addplot+[
dashed, blue, mark options={scale=0},
error bars/.cd, 
    y fixed,
    y dir=both, 
    y explicit
] table [x expr=\thisrow{x}/3600, y=y] {fig/ACAS-noEM.txt};
\addlegendentry{\Large DL, DNN-ACAS Xu w/o \code{Seq\_fresh}}

\addplot+[
dashed, teal, mark options={scale=0},
error bars/.cd, 
    y fixed,
    y dir=both, 
    y explicit
] table [x expr=\thisrow{x}/3600, y=y] {fig/IL-carla-noEM.txt};
\addlegendentry{\Large IL, CARLA w/o \code{Seq\_fresh}}

\addplot+[
dashed, purple, mark options={scale=0},
error bars/.cd, 
    y fixed,
    y dir=both, 
    y explicit
] table [x expr=\thisrow{x}/3600, y=y] {fig/MARL-noEM.txt};
\addlegendentry{\Large MARL, Coop Navi w/o \code{Seq\_fresh}}

\addplot+[
dashed, orange, mark options={scale=0},
error bars/.cd, 
    y fixed,
    y dir=both, 
    y explicit
] table [x expr=\thisrow{x}/3600, y=y] {fig/RL-Game-noEM.txt};
\addlegendentry{\Large RL, BipedalWalker w/o \code{Seq\_fresh}}

\end{axis}
\end{tikzpicture}}
    \vspace{-10pt}
    \caption{\#Crashes found w/ and w/o freshness guidance.}
    \label{fig:crashes}
  \end{figure}

\noindent \textbf{Results.}~As shown in \F~\ref{fig:crashes}, we use the dashed
lines to represent \#crashes without the sequence freshness guidance. That is,
\tool\ is only guided by the cumulative reward (line 18 of
\A~\ref{alg:fuzz-alg}). We observe that with the same initial seed corpus when
\tool\ is not guided by the sequence freshness, \#crashes (dashed lines) is much
smaller than when it is guided by both the cumulative rewards and sequence
freshness (solid lines). We view this comparison as strong evidence to show the
usefulness of sequence freshness.
Note that \F~\ref{fig:crashes} depicts the run with the 
median number of crashes across three separate runs, and 
as indicated in \T~\ref{tab:results}, 
the results of \tool\ are consistent, with a small standard deviation.

When testing DNN-based ACAS Xu, we observe that during the first four hours, the
performance of the two setups is close because the environment of DNN-based ACAS Xu is simpler than
other scenarios, and its model's input dimension is small (only 5). 
However, after four hours, \tool\ can hardly find any new crashes, without the
sequence freshness guidance. It terminates soon since it runs out of seeds
without increasing any cumulative rewards. In contrast, with the sequence
freshness guidance, \tool\ can proceed further and find 37 more crashes in the
following eight hours. Given that the initial seed corpus for both settings is
the same, we deem it efficient to use the state sequence freshness to guide
\tool\ in testing the models of DNN-based ACAS Xu.

\begin{figure}[!t]
    \centering
    \includegraphics[width=1.00\linewidth]{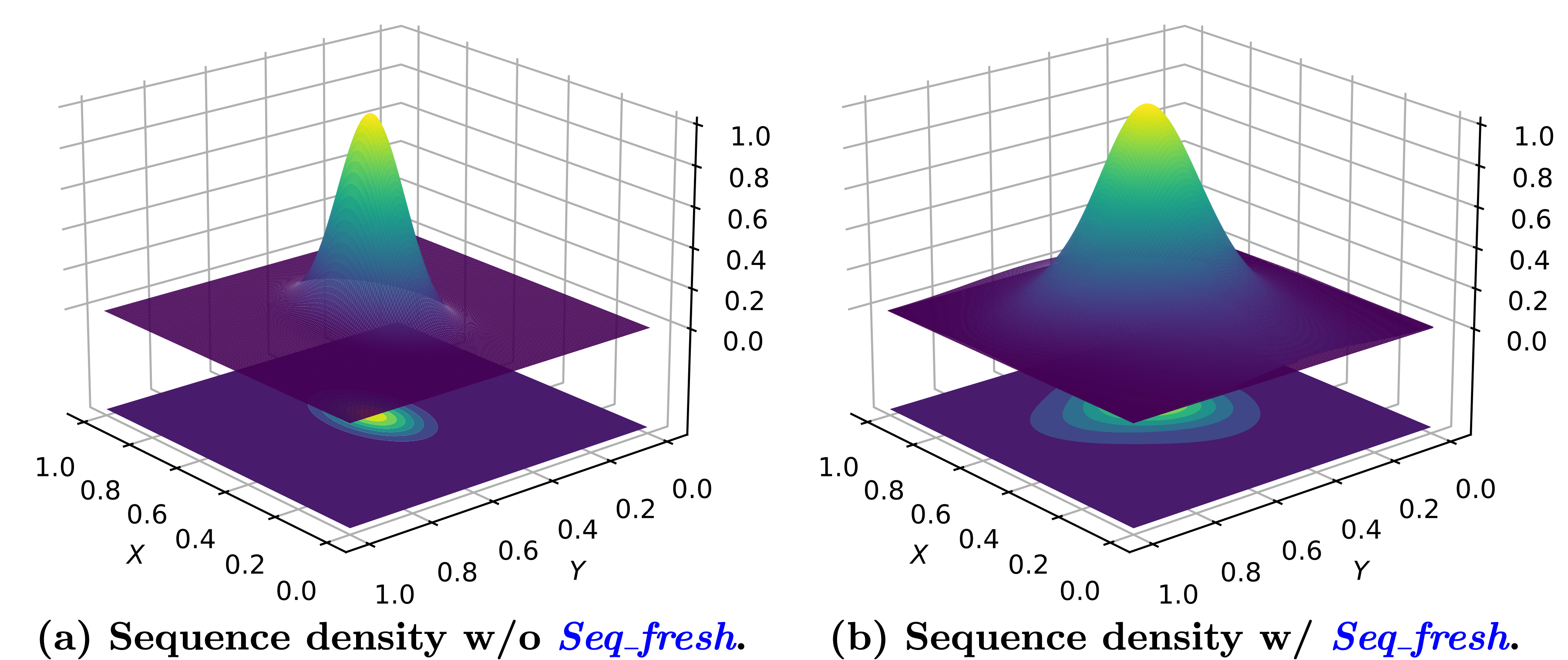}
    \vspace{-10pt}
    \caption{State coverage w/o and w/ freshness guidance.}
    \label{fig:statecvg}
\end{figure}

We use GMMs and DynEM to estimate the distributions of the sequences \tool\ has found, 
whose rationality has been demonstrated in \S~\ref{subsec:design-em}. 
We visualize the GMMs estimated when testing the RL model for CARLA in \F~\ref{fig:statecvg}, where
\F~\hyperref[fig:statecvg]{\ref{fig:statecvg}a} and \F~\hyperref[fig:statecvg]{\ref{fig:statecvg}b} illustrate the estimated distributions for
\tool\ without and with the state sequence freshness guidance, respectively. 
We project the $X$ and $Y$ axes to the same scale to compare the covered areas fairly. 
Comparing the space covered by these two distributions, we conclude that the guidance of state
sequence freshness helps \tool\ cover a larger state sequence space. 
Thus, this guidance can boost the efficiency of \tool\ in
progressively finding diverse state sequences.
GMMs visualization results for other models are at~\cite{snapshot}.

\vspace{-2pt}
\begin{tcolorbox}[size=small]
    \textbf{Answer to RQ2}: The state sequence freshness enhances the efficiency of \tool\ in finding crashes, 
    helps \tool\ explore diverse state sequences more efficiently, 
    and enables \tool\ to cover a larger space of state sequences.
\end{tcolorbox}
\vspace{-2pt}

\subsection{RQ3: Root Cause Analysis}
\label{subsec:evaluation-RQ3}

\begin{figure*}[!t]
    \captionsetup[sub]{font=scriptsize}
    \begin{subfigure}{0.95\textwidth}
        \centering
        \resizebox{1.0\textwidth}{!}{\newenvironment{customlegend}[1][]{%
    \begingroup
    \csname pgfplots@init@cleared@structures\endcsname
    \pgfplotsset{#1}%
}{%
    \csname pgfplots@createlegend\endcsname
    \endgroup
}%

\def\addlegendimage{\csname pgfplots@addlegendimage\endcsname}

\begin{tikzpicture}
\begin{customlegend}[legend columns=3,legend style={align=center,draw=none,column sep=8ex, font=\Large},
        legend entries={\textsc{Neuron activation of CRASH sequences found by \tool},\textsc{Neuron activation of NORMAL sequences}, \textsc{Neuron activation of  RANDOMLY mutated sequences}}]
        \addlegendimage{color=red, mark=*, mark size=3pt, only marks}
        \addlegendimage{color=blue, mark=*, mark size=3pt, only marks}
        \addlegendimage{color=teal, mark=*, mark size=3pt, only marks}
        \end{customlegend}
\end{tikzpicture}}
    \end{subfigure}

        \begin{subfigure}{.185\linewidth}
            \centering
            \resizebox{1.0\linewidth}{!}{
            \begin{tikzpicture}
                \node (img) {\includegraphics{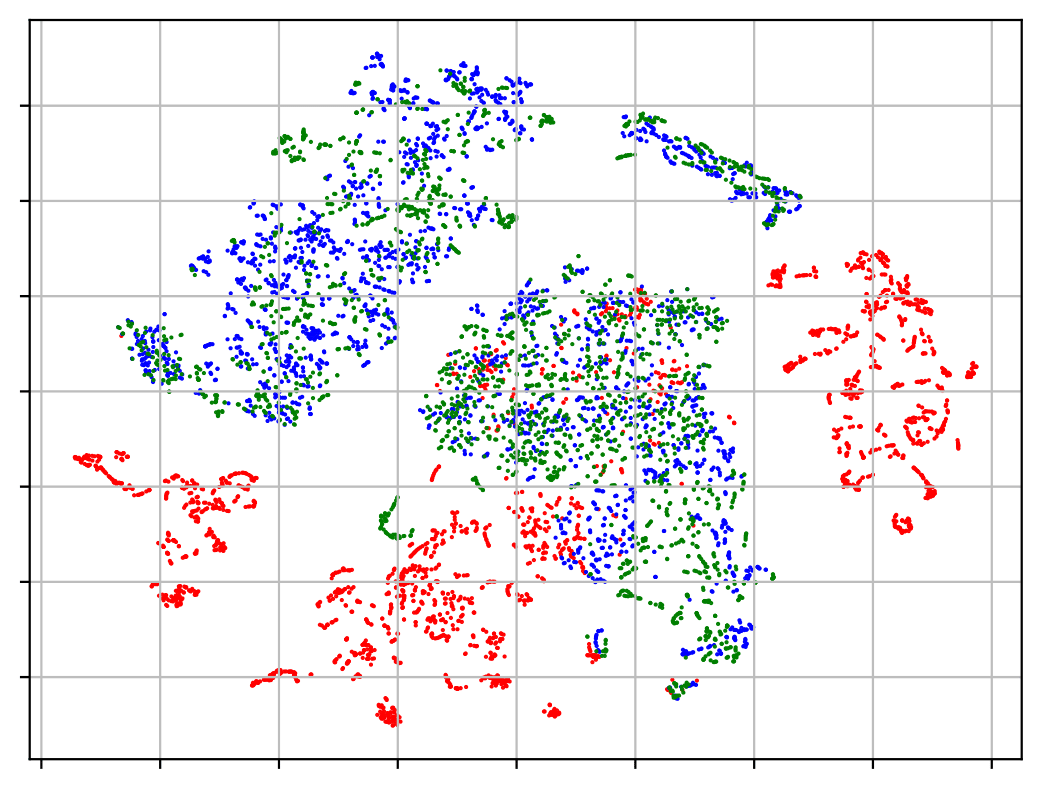}};
            \end{tikzpicture}
            }
            \vskip -5pt
        \caption{RL for CARLA.}
        \end{subfigure}%
        \hspace{1pt}
        \begin{subfigure}{.185\linewidth}
            \centering
            \resizebox{1.0\linewidth}{!}{
            \begin{tikzpicture}
                \node (img) {\includegraphics{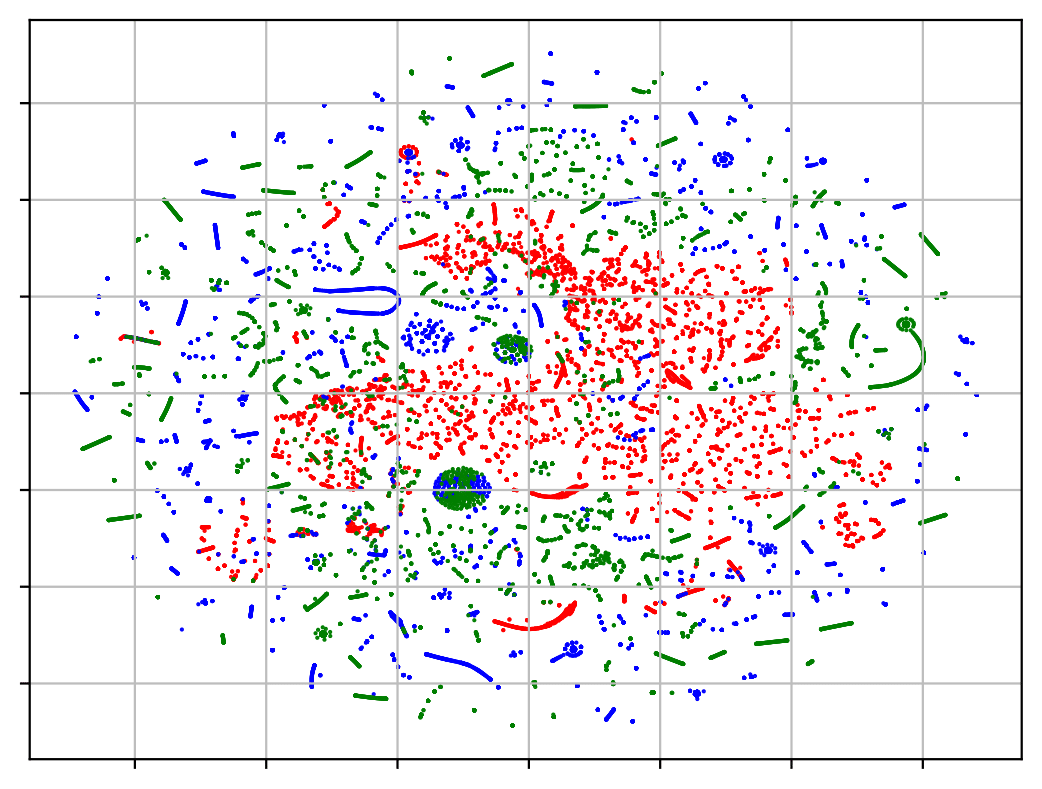}};
            \end{tikzpicture}
            }
            \vskip -5pt
        \caption{DL of DNN-ACAS Xu.}
        \end{subfigure}%
        \hspace{1pt}
        \begin{subfigure}{.185\linewidth}
            \centering
            \resizebox{1.0\linewidth}{!}{
            \begin{tikzpicture}
                \node (img) {\includegraphics{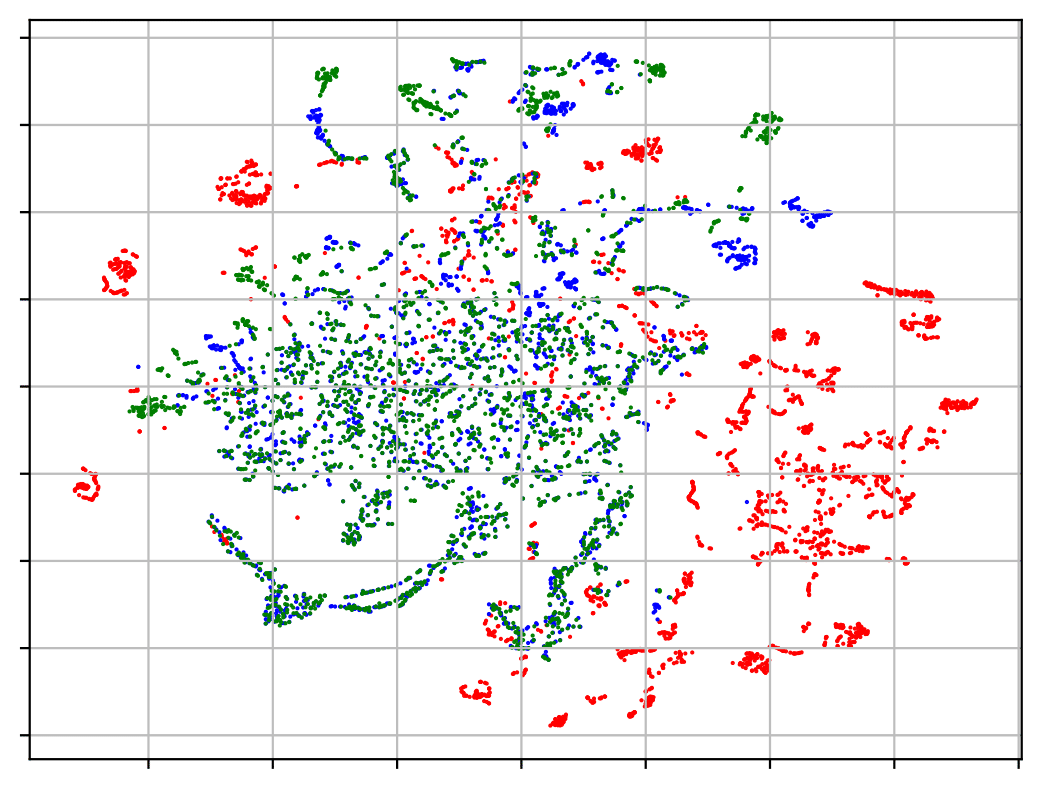}};
            \end{tikzpicture}
            }
            \vskip -5pt
        \caption{IL for CARLA.}
        \end{subfigure}%
        \hspace{1pt}
        \begin{subfigure}{.185\linewidth}
            \centering
            \resizebox{1.0\linewidth}{!}{
            \begin{tikzpicture}
                \node (img) {\includegraphics{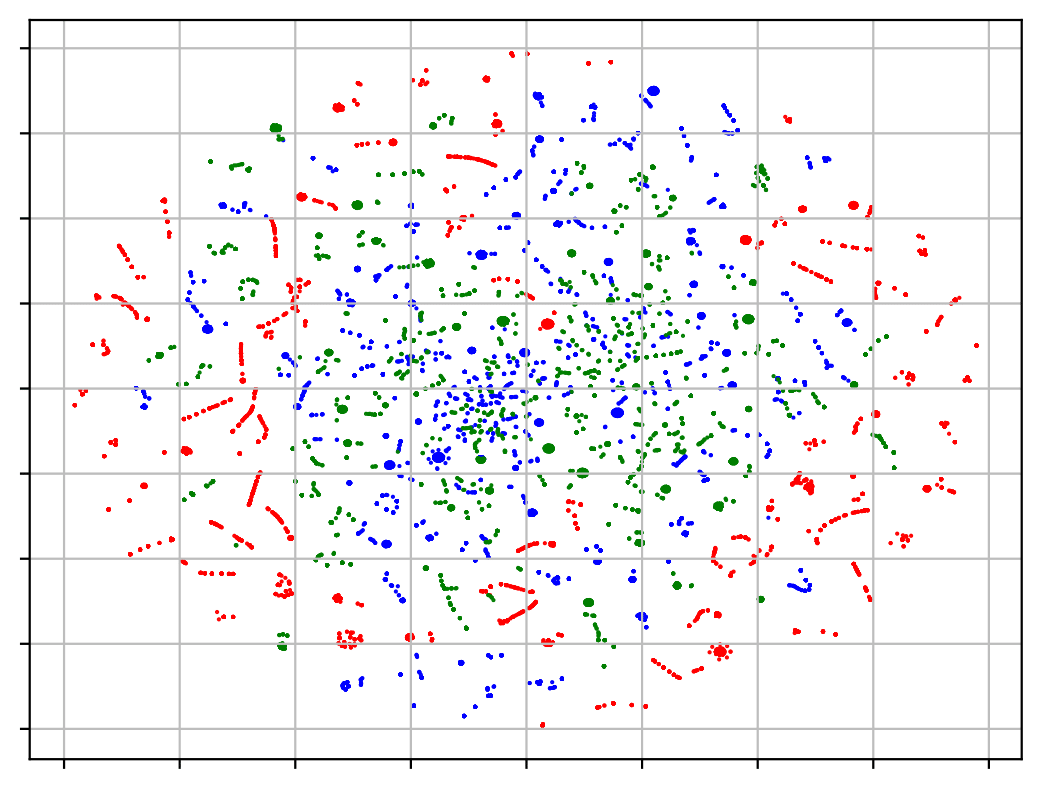}};
            \end{tikzpicture}
            }
            \vskip -5pt
        \caption{MARL for Coop Navi.}
        \end{subfigure}%
        \hspace{1pt}
        \begin{subfigure}{.185\linewidth}
            \centering
            \resizebox{1.0\linewidth}{!}{
            \begin{tikzpicture}
                \node (img) {\includegraphics{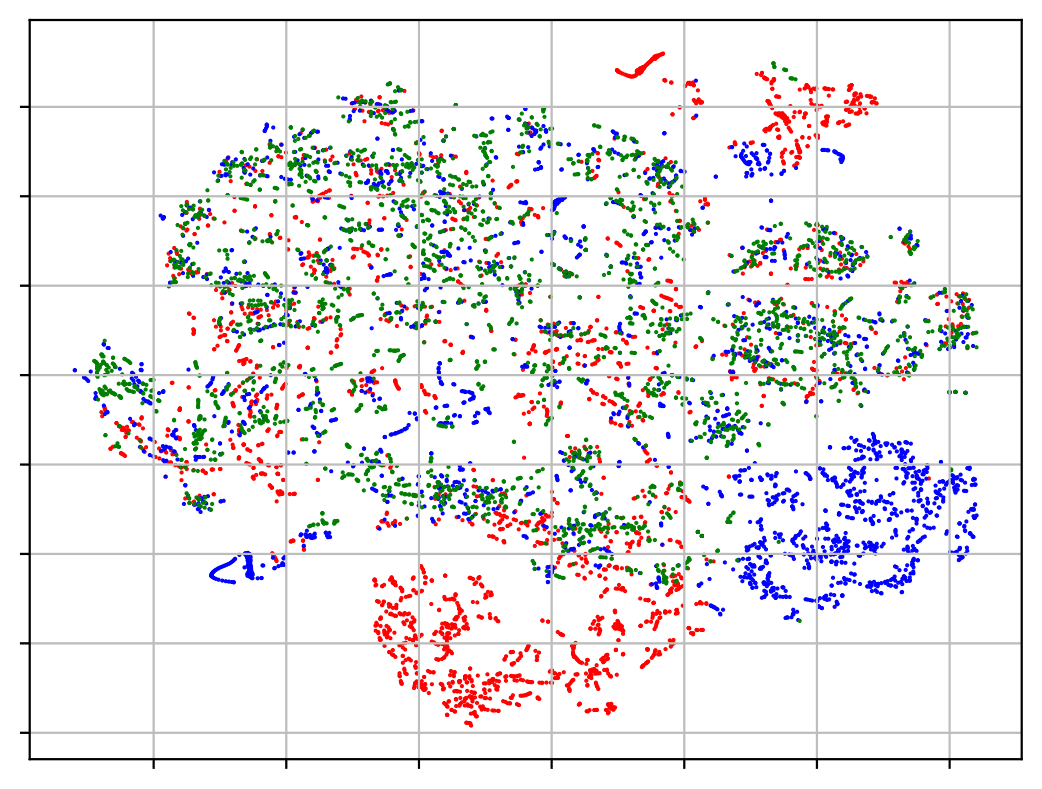}};
            \end{tikzpicture}
            }
            \vskip -5pt
        \caption{RL for BipedalWalker.}
        \end{subfigure}
        \centering
        \vspace{-8pt}
        \caption{Neuron activations projected by t-SNE.}
        \label{fig:root-cause}
        \vspace*{-5pt}
    \end{figure*}

\noindent \textbf{Setup.}~Inspired by contemporary DNN testing
criteria~\cite{xie2019deephunter,pei2017deepxplore,ma2018deepgauge,odena2018tensorfuzz}, 
we characterize crash-triggering states by measuring
their induced neuron activation patterns in all tested models. In this step, we
take a common approach to use t-SNE~\cite{van2008visualizing} to reduce dimensions and
visualize the distributions of activated neurons.

\noindent \textbf{Results.}~\F~\ref{fig:root-cause} visualizes and compares the
states' neuron activations of the target models. We plot the 
activation patterns of the states in sequences when their initial states are crash-triggering (\textcolor{red}{red dots}), normal (\textcolor{blue}{blue
  dots}), and randomly-mutated (\textcolor{teal}{teal dots}).
The neuron activations are projected to two dimensions by t-SNE.

The neuron activations of crash-triggering
state sequences found by \tool\ have a clear boundary with normal and randomly
mutated sequences, which shows that the crash-triggering states can 
trigger the models' abnormal internal logics. 
The results are promisingly consistent across all models of different paradigms. 
Furthermore, randomly mutated states are mostly mixed with normal states, indicating that 
random mutation with no guidance can hardly provoke corner case
behaviors of the models solving MDPs.

In \S~\ref{sec:implementation}, we have clarified that the mutated initial
states are validated and solvable: crashes are avoidable if the tested models
can take optimal actions (therefore not buggy). Thus, we emphasize that the
crashes found by \tool\ are not due to unrealistic states. Rather, they share
similar visual appearances with normal states and can occur in real-world
scenarios. Viewing that real-world models that solve MDPs might be under high
chance of being ``vulnerable'' toward these stealthy states found by \tool, our
findings regarding their distinct neuron activation patterns is inspiring. In
particular, we envision high feasibility for hardening real-world models solving
MDPs by detecting abnormal model logics according to their neuron activation
patterns; see \S~\ref{subsec:evaluation-RQ4}.

\vspace{-2pt}
\begin{tcolorbox}[size=small]
    \textbf{Answer to RQ3}: Though crash-triggering initial states are
    considered \textit{normal} from the MDP environments' perspective, they
    induce distinct neuron activation patterns. This shows that \tool\ can cover
    the model's corner internal logics and trigger its abnormal behaviors more
    efficiently than randomly-mutated state sequences. This finding also reveals
    exciting potentials of detecting crash-triggering states.
\end{tcolorbox}
\vspace{-2pt}

\subsection{RQ4: Enhance Model Robustness}
\label{subsec:evaluation-RQ4}

\noindent \textbf{Setup.}~The findings in \S~\ref{subsec:evaluation-RQ3} inspire 
us to develop a cluster to detect the models' abnormal internal logics. 
We use the Mean-Shift clustering technique~\cite{cheng1995mean} to distinguish 
between the models' normal and abnormal behaviors automatically. More specifically,
we first calculate the clustering centers of normal and abnormal neuron activations. 
When a new state's model activation is observed, we measure its distance between the normal
and abnormal clusters. If it is too far from the normal clusters, we then regard it as abnormal
behavior. The size of the dataset we use for clustering is 6,000, half of which are abnormal
neuron activations found by \tool. We then randomly split 20\% of the entire dataset as test data to assess 
the performance of our detector.

In addition, we repair the models used by DNN-based ACAS Xu with the findings of
\tool. Here, ``repair'' represents a standard data augmentation
procedure~\cite{feng2020genaug, yang2019data, gleave2019adversarial}, where we
construct a dataset to fine tune the model. Our fine-tuning dataset includes the
crash-triggering sequences found by \tool\ and randomly sampled sequences, both
of which contain 13,600 frames.
We manually label the crash-triggering frames with the optimal actions that can
avoid collisions. Then, we re-run fuzz testing with the same settings as in
\S~\ref{subsec:evaluation-RQ1} to assess our repairment. Further, we randomly
select 3,000 initial states and compare their cumulative rewards before and after the models have been repaired to measure
the performance of the repaired models on normal cases. We underline that other
models can be fine-tuned and made more robust in the same way. Because of its
widespread usage in safety-critical circumstances and its relatively simple
architecture, we chose DNN-based ACAS Xu as a proof of concept. Training complex models
solving MDPs requires considerable computation resources, e.g., training the
RL model for CARLA takes 57 days~\cite{toromanoff2020end}.

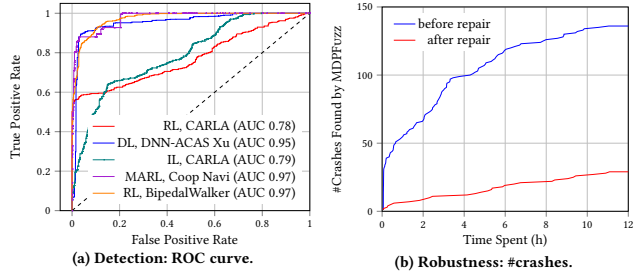
\begin{figure}[!t]
  \captionsetup[sub]{font=scriptsize}
    \begin{subfigure}{0.49\linewidth}
		\centering
		\resizebox{1.0\linewidth}{!}{\begin{tikzpicture}
    \begin{axis}[
      grid=major,
      legend cell align={right},
      xmin=-0.05, xmax=1, ymin=0, ymax=1.05,
      ytick align=outside, ytick pos=left,
      xtick align=outside, xtick pos=left,
      xlabel={\Large False Positive Rate},
      ylabel={\Large True Positive Rate},
      legend pos=south east,
      legend style={draw=none}]

\addplot+[
red, mark options={scale=0},
error bars/.cd, 
    y fixed,
    y dir=both, 
    y explicit
] table [x=x, y=y] {fig/roc-RLcarla.txt};
\addlegendentry{\Large RL, CARLA (AUC 0.78)}

\addplot+[
  blue, mark options={scale=0},
  error bars/.cd, 
      y fixed,
      y dir=both, 
      y explicit
  ] table [x=x, y=y] {fig/roc-ACAS.txt};
\addlegendentry{\Large DL, DNN-ACAS Xu (AUC 0.95)}

\addplot+[
  teal, mark options={scale=0},
  error bars/.cd, 
      y fixed,
      y dir=both, 
      y explicit
  ] table [x expr=1-\thisrow{x}, y expr=1-\thisrow{y}] {fig/roc-IL.txt};
\addlegendentry{\Large IL, CARLA (AUC 0.79)}

\addplot+[
  purple, mark options={scale=0},
  error bars/.cd, 
      y fixed,
      y dir=both, 
      y explicit
  ] table [x=x, y=y] {fig/roc-MARL.txt};
\addlegendentry{\Large MARL, Coop Navi (AUC 0.97)}

\addplot+[
  orange, mark options={scale=0},
  error bars/.cd, 
      y fixed,
      y dir=both, 
      y explicit
  ] table [x=x, y=y] {fig/roc-RL.txt};
\addlegendentry{\Large RL, BipedalWalker (AUC 0.97)}

\addplot+[
dashed, black, mark options={scale=0},
error bars/.cd, 
    y fixed,
    y dir=both, 
    y explicit
] table [x=x, y=y] {fig/roc-reference.txt};

\end{axis}
\end{tikzpicture}}
    \vskip -5pt
        \caption{Detection: ROC curve.}
	\label{fig:repair-roc}
	\end{subfigure}%
  \hspace{1pt}
  \begin{subfigure}{0.49\linewidth}
		\centering 
    \resizebox{1.0\linewidth}{!}{\begin{tikzpicture}
    \begin{axis}[
      grid=major,
      legend cell align={right},
      xmin=0, xmax=12, ymin=0, ymax=150,
      ytick align=outside, ytick pos=left,
      xtick align=outside, xtick pos=left,
      xlabel={\Large Time Spent (h)},
      ylabel={\Large \#Crashes Found by \tool},
      legend pos=north west,
      legend style={draw=none}]

\addplot+[
  blue, mark options={scale=0},
  error bars/.cd, 
      y fixed,
      y dir=both, 
      y explicit
  ] table [x expr=\thisrow{x}/3600, y=y] {fig/beforerepair.txt};
  \addlegendentry{\Large before repair}

\addplot+[
red, mark options={scale=0},
error bars/.cd, 
    y fixed,
    y dir=both, 
    y explicit
] table [x expr=\thisrow{x}/3600, y=y] {fig/afterrepair.txt};
\addlegendentry{\Large after repair}

\end{axis}
\end{tikzpicture}}
    \vskip -5pt
        \caption{Robustness: \#crashes.}
	\label{fig:repair-crash}
	\end{subfigure}
  \vspace{-8pt}
    \caption{Enhancing the models' robustness.}
	\label{fig:repair}
%  \vspace{-10pt}
\end{figure}

\noindent \textbf{Results.}~In \F~\ref{fig:repair-roc}, we report the performance of our detector on test data. 
The area under the receiver operating characteristic curve (AUC-ROC) is above 0.78 for each model, indicating that our detector can simultaneously achieve good precision and recall in detecting abnormal model logics. 
The AUC-ROC is a widely-used metric for binary classifiers; a larger AUC-ROC indicates a better performance of the detector. 
In \F~\ref{fig:repair-crash}, we present the results of comparing the models in DNN-based ACAS Xu before and after repairing, which show that the \#crash detected by \tool\ in 12 hours after model repairing is substantially lower, at 29, than 139 before repairing. 
Thus, after fine-tuning, the model with fewer crashes detected by \tool becomes more robust.
Furthermore, on the 3,000 randomly-selected normal sequences (which
are not crash-triggering), 
the average cumulative rewards of the models before and after repair are close, which are $30.73$ and $30.74$, respectively.
The results reveal that the models after repair can still perform well under normal states.

\vspace{-4pt}
\begin{tcolorbox}[size=small]
  \textbf{Answer to RQ4}: Our detector can detect models' abnormal internal
  logics, enhancing the models' robustness and avoiding catastrophic crashes.
  Also, with model repairing, findings of \tool\ substantially contribute to the
  models' robustness without affecting, if not boosting, their accuracy.
\end{tcolorbox}
\vspace{-4pt}

\section{Discussion}
\label{sec:discussion}

\noindent \textbf{Threat To Validity and Limitations.}~Construct validity
represents the extent to which \tool\ actually reflects the correctness of MDP
models. Overall, \tool\ launches dynamic testing to study MDP models. One threat
is that \tool\ cannot find all crash-triggering inputs, let alone ensure the
functional correctness of MDP models. \tool\ roots the same goal as most
previous works (\S~\ref{sec:related}) to test models rather than verify their
correctness. On the other hand, we deem that \tool\ is an effective testing
tool, given the large search space and the fact that \tool\ targets different
(blackbox) models solving MDPs. 
We design \tool\ to address potential bias in mutations and their
representativeness of real bugs. Particularly, when mutating inputs, \tool\ uses
sequence freshness as the feedback to guide mutation. Therefore, \tool\
progressively explores the sequence search space, reducing the bias of being
trapped within local regions. We add random noise to only the initial states,
and all mutated initial states can pass the validation procedures of MDP
environments. Thus, the input states and the induced sequences are realistic. To
mitigate threats of using only ``random noise,'' it is interesting to consider
``semantics-level'' mutations (e.g., changing roads, buildings). However, that
may be difficult, forcing us to manually prepare some templates for mutation and
thereby undermining generalization. 

Another threat is that bugs may be potentially overlooked since \tool\ only
mutates the entry states (\S~\ref{sec:design}). Nevertheless, as discussed in
\S~\ref{sec:motivation}, mutating multiple middle states in a sequence can
presumably generate unrealistic inputs due to the continuity of adjacent states.
We deem it an interesting future work to generate realistic ``mutated paths''
with low cost. 

We also consider mitigating potential biases in the abnormal behavior detector.
To train the abnormal behavior detector, we use the same amount of abnormal data
found by \tool\ and random sampled normal data. As shown in \F~\ref{fig:repair},
our detector has high TPR and very low FPR, indicating that it is not biased. 

\noindent \textbf{State Sequence Coverage.}~To test models solving MDPs, \tool\
implicitly increases state sequence coverage by estimating freshness. Readers
may wonder whether \tool\ can leverage existing DNN coverage criteria, e.g.,
neuron coverage~\cite{pei2017deepxplore} or surprise
adequacy~\cite{kim2019guiding, kim2020reducing}. We clarify that it is 
challenging. Unlike previous works~\cite{pei2017deepxplore, ma2018deepgauge,
xie2019deephunter, kim2019guiding, kim2020reducing}, it's hard to discretize the
state sequence considering its large space. Second, state sequences can be
arbitrarily long, whereas \#neurons used to form prior coverage criteria are
generally fixed. Moreover, existing criteria are designed to assess 
quality (``surprise'') of \textit{each} DNN input, whereas we assess 
freshness of one state sequence collected when models solving MDPs respond to
\textit{a series} of inputs. Our current design consider relation/dependency
among different states in a sequence, whereas prior works generally treat
different inputs separately. Plus, \tool\ is designed to test models in
\textit{blackbox} settings, where existing coverage criteria are generally
designed for \textit{whitebox} settings. We leave it as future work
to propose coverage criteria in high-dimensional space and consider dependency
among dimensions.

\noindent \textbf{Comparison with DynPCA.}~There exist metrics for measuring the
similarity of MDP states~\cite{ferns2012methods,ferns2011bisimulation,
  castro2020scalable}. As clarified in \S~\ref{subsec:design-em}, comparing the 
newly-discovered state sequence with each historial state sequence
is too costly. DynPCA~\cite{manes2020ankou}, as an online version of PCA,
calculates the Euclidean distance between the 
new seed and prior seeds in a
smaller latent space. Our method, \textsc{DynEM}, shares similar concepts with DynPCA.
However, DynPCA is not directly compatible with \tool.

First, PCA captures only uncorrelated components by assuming a 
constant multivariate Gaussian distribution for variables~\cite{rummel1988applied}.
However, the variables of states in MDPs are usually continuous, and their
principal components can be highly nonlinear, making standard PCA 
useless. Considering a sequence of 100 frames where each state in the frame has
64 dimensions, the sequence has 6,400 continuous variables. No multivariate
Gaussian distribution can be guaranteed, especially as states are not
independent. Using the Markov property clarified in \S~\ref{subsec:design-em},
\tool\ only needs to estimate two distributions to estimate an arbitrarily long state sequence. Further, GMMs can estimate any
smooth distribution, not just multivariate Gaussian, as stated in
\S~\ref{subsec:design-em}.
Second, DynPCA fixes the input dimensions; hence in MDPs, the sequence length
must be fixed. In our method, as stated in \S~\ref{sec:implementation}, the
sequence length can be changed even during the same
fuzzing campaign. For instance, we can prolong the sequence length to better
stress the target model in case it is shown as robust during online fuzzing.

\section{Conclusion}

We present \tool, a blackbox testing tool for models solving MDPs. \tool\
detects whether the models enter abnormal and dangerous states. \tool\
incorporates optimizations to efficiently fuzz models of different paradigms and
under different scenarios. We show that crash-triggering states result in
distinct neuron activation patterns. Based on these findings, we harden the
tested models using an abnormal internal logics detector. We repair models using
\tool's findings to enhance their robustness without sacrificing accuracy.

\bibliographystyle{ACM-Reference-Format}
\bibliography{bib/main}

\end{document}